\pdfoutput=1

\documentclass[11pt,twoside,a4paper,cmspaper,final,collab]{cms-tdr}
\def\svnVersion{340935}\def\cmsCernNoTag{CERN-PH-EP/2015-332}\def\cmsCernDate{\today}\def\cmsMessage{Published in the European Physical Journal C as \href{http://dx.doi.org/10.1140/epjc/s10052-016-4067-z}{\doi{10.1140/epjc/s10052-016-4067-z.}}}

\begin{document}\cmsNoteHeader{EXO-14-010}

\hyphenation{had-ron-i-za-tion}
\hyphenation{cal-or-i-me-ter}
\hyphenation{de-vices}
\RCS$Revision: 331769 $
\RCS$HeadURL: svn+ssh://svn.cern.ch/reps/tdr2/papers/EXO-14-010/trunk/EXO-14-010.tex $
\RCS$Id: EXO-14-010.tex 331769 2016-03-10 08:45:13Z qili $
\newlength\cmsFigWidth
\ifthenelse{\boolean{cms@external}}{\setlength\cmsFigWidth{0.48\textwidth}}{\setlength\cmsFigWidth{0.7\textwidth}}
\ifthenelse{\boolean{cms@external}}{\providecommand{\cmsLeft}{top\xspace}}{\providecommand{\cmsLeft}{left\xspace}}
\ifthenelse{\boolean{cms@external}}{\providecommand{\cmsRight}{bottom\xspace}}{\providecommand{\cmsRight}{right\xspace}}
\ifthenelse{\boolean{cms@external}}{\providecommand{\cmsTopLeft}{top\xspace}}{\providecommand{\cmsTopLeft}{upper left\xspace}}
\ifthenelse{\boolean{cms@external}}{\providecommand{\cmsTopRight}{middle\xspace}}{\providecommand{\cmsTopRight}{upper right\xspace}}

\providecommand{\CLs}{\ensuremath{CL_\mathrm{s}}\xspace}
\newcommand{\THISLUMI} {19.7\xspace}
\newcommand{\MINmWHMASS} {700\xspace}
\newcommand{\MAXmWHMASS} {2500\xspace}
\newcommand{\MINmWHMASSCONCL} {0.8\xspace}
\newcommand{\MAXmWHMASSCONCL} {2.5\xspace}
\newcommand{\SFTTBARELEHP} {\ensuremath{0.96\pm0.03}\xspace}
\newcommand{\SFTTBARMUHP} {\ensuremath{0.97\pm0.02}\xspace}
\newcommand{\SFTTBARELELP} {\ensuremath{1.39\pm0.08}\xspace}
\newcommand{\SFTTBARMULP} {\ensuremath{1.31\pm0.05}\xspace}
\newcommand{\SFWTAGHP} {\ensuremath{0.89\pm0.08}\xspace}
\newcommand{\SFWTAGLP} {\ensuremath{1.28\pm0.30}\xspace}
\newcommand{\WMASSDATA} {\ensuremath{84.7\pm0.4}\xspace}
\newcommand{\WMASSMC} {\ensuremath{83.4\pm0.3}\xspace}
\newcommand{\WRESDATA} {\ensuremath{7.9\pm0.6}\xspace}
\newcommand{\WRESMC} {\ensuremath{7.2\pm0.4}\xspace}
\newcommand{\WJETNORMUNCERT} {below 40\%\xspace}
\newcommand{\TTBARNORMUNCERT} {5.6\%\xspace}
\newcommand{\VVNORMUNCERT} {10\%\xspace}
\newcommand{\VTAGUNCERTHP} {9\%\xspace}
\newcommand{\VTAGUNCERTLP} {24\%\xspace}
\newcommand{\LUMIUNCERT} {2.6\xspace}
\newcommand{\MEANUNCERT} {1.5\%\xspace}
\newcommand{\WIDTHUNCERT} {4.5\%\xspace}
\newcommand{\RANGELIMITS} {from 70\unit{fb} to 3\unit{fb}\xspace}

\newcommand{\pp}{\Pp\Pp}%
\newcommand{\Wo}{\PW\xspace}%
\newcommand{\Ho}{\PH\xspace}%
\newcommand{\Wp}{\PWp\xspace}%
\newcommand{\Wm}{\PWm\xspace}%
\newcommand{\Zo}{\cPZ\xspace}%
\newcommand{\Vo}{\ensuremath{\mathrm{V}}\xspace}%
\newcommand{\MN}{\Pgm\Pgm\xspace}%
\newcommand{\EN}{\Pe\Pgn\xspace}%
\newcommand{\LN}{\ensuremath{\ell\Pgn}\xspace}%
\newcommand{\MW}{\ensuremath{m_\Wo}\xspace}%
\newcommand{\MZ}{\ensuremath{m_\Zo}\xspace}%
\providecommand{\MT}{\ensuremath{M_\mathrm{T}}\xspace}%
\newcommand{\MLL}{\ensuremath{m_{\ell\ell}}\xspace}%
\newcommand{\nunubar}{\Pgn\Pagn\xspace}%
\newcommand{\qqbarpr}{\ensuremath{\Pq\Paq^({}'^){}}\xspace}
\newcommand{\HZZllqq}{\ensuremath{\HZZ\to\qqbar\,\LL}}%
\newcommand{\MPl}{\ensuremath{{M_{\text{Pl}}}}\xspace}%
\newcommand{\RedMPl}{\ensuremath{\overline{M}_{\text{Pl}}}\xspace}%
\newcommand{\GZZllqq}{\ensuremath{\GZZ\to\qqbar\,\LL}}%
\newcommand{\Gllqq}{\ensuremath{\Grav\to\LL\qqbar}}%
\newcommand{\XWH}{\ensuremath{X\to\Wo\Ho}}%
\newcommand{\Xllqq}{\ensuremath{X\to\LL\qqbar}}%
\newcommand{\XZZllqq}{\ensuremath{\XZZ\to\LL\qqbar}}%
\newcommand{\Mg}{\ensuremath{m_{\PWpr}}}%
\newcommand{\mG}{\ensuremath{m_{\PWpr}}}%
\newcommand{\mX}{\ensuremath{\text{M}_{\text{X}}}}%
\newcommand{\wX}{\ensuremath{\Gamma_{\text{X}}}}%
\newcommand{\mZ}{\ensuremath{m_{\Zo}}}%
\newcommand{\mW}{\ensuremath{m_{\Wo}}}%
\newcommand{\mWH}{\ensuremath{\mbox{M}_{\Wo\Ho}}\xspace}%
\newcommand{\mll}{\ensuremath{m_{\ell\ell}}\xspace}%
\newcommand{\mLL}{\ensuremath{m_{\ell\ell}}\xspace}%
\newcommand{\mjj}{\ensuremath{m_\mathrm{jj}}\xspace}%
\newcommand{\mJ}{\ensuremath{m_{\text{jet}}}}%
\newcommand{\ptj}{\ensuremath{{p_{\text{T}}^{\text{jet}}}}}
\newcommand{\nsubj}{\ensuremath{\tau_{21}}}%
\newcommand{\ktilde}{\ensuremath{k/\overline{M}_\mathrm{Pl}}\xspace}%
\newcommand{\JHUGEN} {{\textsc{jhugen}}\xspace}
\newcommand{\lnujet}{\ensuremath{\ell \nu}+V\text{-jet}\xspace}
\newcommand{\lljet}{\ensuremath{\ell \ell}+V\text{-jet}\xspace}
\newcommand\T{\relax}

\cmsNoteHeader{EXO-14-010}
\title{Search for massive WH resonances decaying into the $\ell \nu\bbbar$ final state at $\sqrt{s}=8$\TeV }

\date{\today}

\abstract{
A search for a massive resonance \PWpr decaying into a W and a Higgs boson in the $\ell \nu \bbbar$ ($\ell = \Pe$, $\mu$) final state is presented.
Results are based on data corresponding to an integrated luminosity of 19.7\fbinv of proton-proton
collisions at $\sqrt{s}=8$\TeV, collected using the CMS detector at the LHC.
For a high-mass ($\gtrsim$1\TeV) resonance, the two bottom quarks coming
from the Higgs boson decay are reconstructed as a single jet, which can be tagged by placing requirements on its substructure and flavour.
Exclusion limits at 95\% confidence level are set on the production cross section of a narrow resonance decaying into WH, as a function of its mass.
In the context of a little Higgs model, a lower limit on the $\PWpr$ mass of 1.4\TeV is set.
In a heavy vector triplet model that mimics the properties of composite Higgs models, a lower limit on the $\PWpr$ mass of 1.5\TeV is set. In the context of this model, the results are combined with related searches to obtain a lower limit on the $\PWpr$ mass of 1.8\TeV, the most restrictive to date for decays to a pair of standard model bosons.}

\hypersetup{%
pdfauthor={CMS Collaboration},%
pdftitle={Search for massive WH resonances decaying into the l nu b anti-b final state at sqrt(s)=8 TeV},%
pdfsubject={CMS},%
pdfkeywords={CMS, physics, W', exotic bosons}}

\maketitle

\section{Introduction}
\label{sec:intro}
This paper presents a search for massive resonances decaying into a W and
a standard model (SM) Higgs boson (H)~\cite{Chatrchyan201230,ATLASHiggsDiscovery,CMSHiggsDiscoveryLong,Aad:2015zhl} in the
$\ell \nu {\rm b \bar{b}}$ ($\ell = \Pe$, $\mu$) final state. Such processes are distinctive features of several extensions of the SM
such as composite Higgs~\cite{Composite0,Composite1,Composite2}, SU(5)/SO(5) Littlest Higgs (LH)~\cite{Han:2003wu,Perelstein:2005ka,Schmaltz:2005ky,Arkani:2002LH},
technicolor~\cite{Lane:2014vca,Lane:2015fza}, and left-right symmetric models~\cite{Dobrescu:2015yba}.
These models provide solutions to the hierarchy problem and
predict new particles including additional gauge bosons such as a heavy $\PWpr$.
The $\PWpr$ in these models can have large branching fractions to WH and WZ,
while the decays to fermions can be suppressed.
The recently proposed heavy vector triplet (HVT) model~\cite{Pappadopulo:2014qza}
generalizes a large class of specific models that predict new heavy spin-1 vector bosons.
In this model, the resonance is described by a simplified Lagrangian in terms of a small number of parameters
representing its mass and couplings to SM bosons and fermions.

For a $\PWpr$ with SM couplings to fermions and thus reduced decay branching ratio to SM bosons,
the most stringent limits on production cross sections are reported in searches with
leptonic final states~\cite{Khachatryan:2014tva,ATLASwprimePAPER2014}. The current lower
limit on the $\PWpr$ mass is 3.3\TeV.
In the same context, searches for a $\PWpr$ decaying into a pair of SM vector bosons (WZ)~\cite{EXO-12-024,CMSwprimeWZPAS8TeV,ATLASWWPAPER8TeV,ATLASwprimeWZPAS}
provide a lower mass limit of 1.7\TeV.
In the context of a HVT model with reduced couplings to fermions (HVT model B), the most stringent limit of 1.7\TeV on the $\PWpr$/$\PZpr$ mass is set by a search for ${\rm W^\prime/Z^\prime \to WH/ZH \to q\bar{q}b\bar{b}}$~\cite{Khachatryan:2015bma}.
The same model is used to interpret the results of a search for ${\rm W^\prime/Z^\prime \to WH/ZH} \to \ell\nu /\ell\ell/\nu\nu+{\rm b\bar{b}}$~\cite{altas-VH-dijet}. A lower limit on the $\PWpr$ mass of 1.5\TeV is set in the same final state reported in Ref.~\cite{altas-VH-dijet}.
Finally, a specific search for ${\rm Z^\prime \to ZH \to q\bar{q}}\tau^{+}\tau^{-}$
was reported in Ref.~\cite{cms-HZ-tautaujet} and interpreted in the context of the same HVT model B.

\begin{figure}[!h]
\begin{center}
\includegraphics[width=0.43\textwidth]{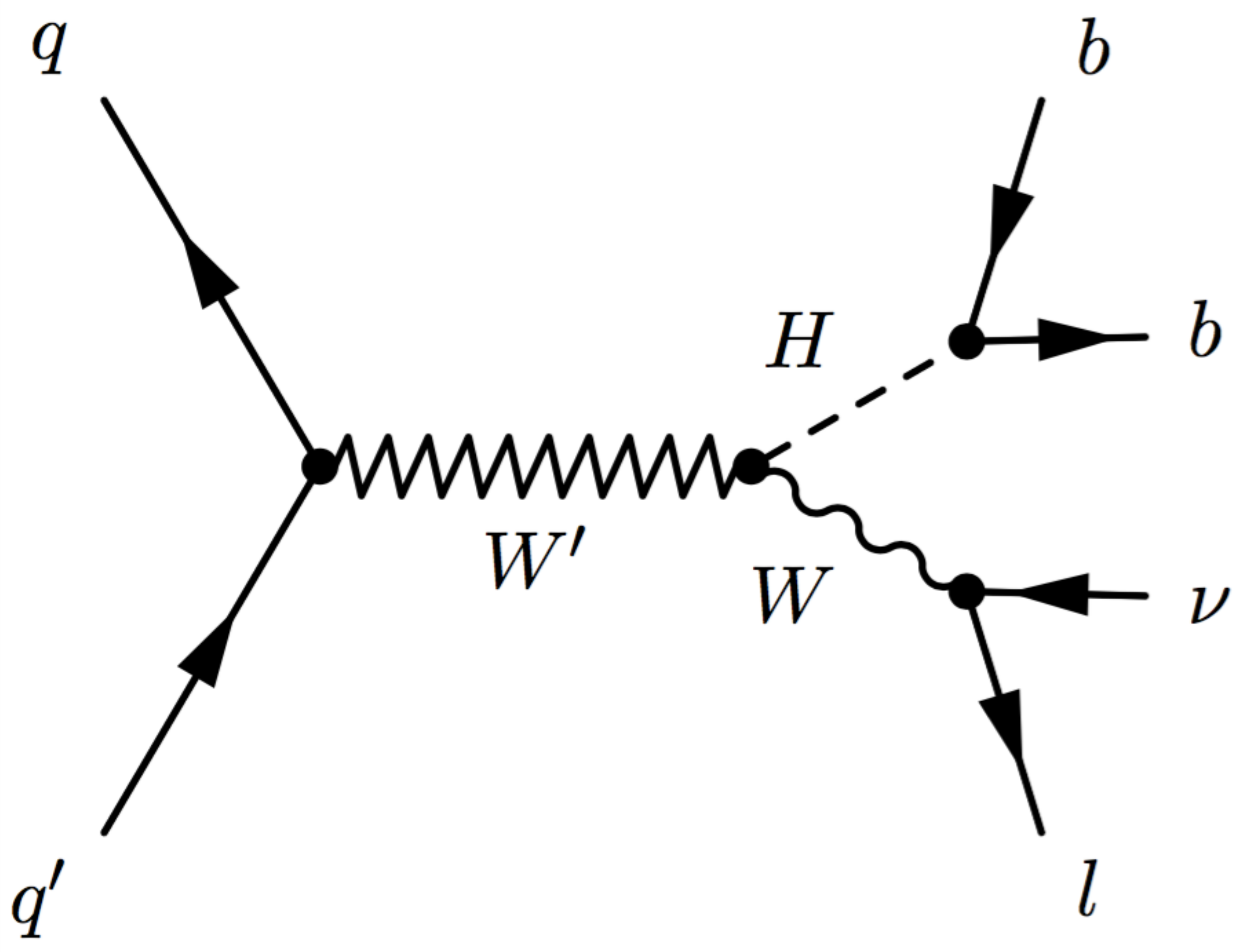}
\end{center}
\caption{Production of a resonance decaying into WH.}\label{fig:feynman}
\end{figure}

This analysis is based on proton-proton collision data at $\sqrt{s}=8$\,TeV collected by the
CMS experiment at the CERN LHC during 2012, corresponding to
an integrated luminosity of 19.7\fbinv.
The signal considered is the production of a resonance with mass above 0.8\TeV
decaying into WH, where the Higgs boson decays into a bottom quark-antiquark pair and the W boson decays into a charged lepton and
a neutrino (Fig.~\ref{fig:feynman}). It is assumed that the resonance is narrow, i.e. that its intrinsic
width is much smaller than the experimental resolution.

The search strategy is closely related to the search for high mass WW resonances
in the $\ell \nu \rm q \bar{q}$ final state, described in Ref.~\cite{EXO-13-009}, with the addition of b tagging techniques.
We search for resonances in the invariant mass of the WH system on top of a smoothly falling background distribution, where the background mainly comprises
events involving pair produced top quarks (\ttbar) or a W boson produced in association with jets (W+jets).
For the resonance mass range considered, the two quarks from the Higgs boson decay would be separated by a small angle, resulting in the detection of a single jet after hadronization.
This jet is tagged as coming from a Higgs boson
through the estimation of its invariant mass, application of jet substructure techniques~\cite{JME-13-006}, and use of specialized b tagging techniques for high transverse momentum ($\pt$) Higgs bosons~\cite{CMS:BTV13001}. 

The results of this analysis are also combined with two previous results~\cite{Khachatryan:2015bma,cms-HZ-tautaujet} to obtain a further improvement in sensitivity.

\section{CMS detector}

The central feature of the CMS apparatus is a superconducting
solenoid of 6\unit{m} internal diameter, providing a field of 3.8\unit{T}. Within the field volume are a silicon
pixel and strip tracker, a crystal electromagnetic calorimeter (ECAL), and a brass and scintillator
hadronic calorimeter (HCAL).
The CMS tracker consists of 1440 silicon pixel and 15\,148 silicon strip
detector modules covering a pseudorapidity range of $\abs{\eta}< 2.5$.
The ECAL consists of nearly 76\,000 lead tungstate crystals, which provide
coverage of $\abs{\eta}< 1.48$ in the central barrel region and $1.48 <\abs{\eta} < 3.00$
in the two forward endcap regions. The HCAL consists of a sampling calorimeter~\cite{Chatrchyan:2008zzk}, which utilizes alternating layers
of brass as an absorber and plastic scintillator as an active material, covering the range $\abs{\eta}< 3$, and is extended to $\abs{\eta}< 5$ by a forward hadron calorimeter.
Muons are measured in the range $\abs{\eta}< 2.4$ with detection planes which employ three technologies: drift tubes, cathode strip chambers, and resistive-plate chambers.
The muon trigger combines the information from the three sub-detectors with a coverage up to $\abs{\eta}<2.1$.
A more detailed description of the CMS detector, together with a definition of the coordinate system used
and the relevant kinematic variables, can be found in Ref.~\cite{Chatrchyan:2008zzk}.

\section{Simulated samples}

For the modelling of the background we use the \MADGRAPH v5.1.3.30~\cite{Alwall:2011uj} event generator to simulate
the production of W boson and Drell--Yan events in association with jets, the \POWHEG
1.0 r1380~\cite{Nason:2004rx,Frixione:2007vw,Alioli:2010xd,Alioli:2009je,Re:2010bp,Alioli:2011as}
package to generate \ttbar and single top quark events, and \PYTHIA v6.424 \cite{Sjostrand:2006za} for diboson
(WW, WZ, and ZZ) processes. All simulated event samples are generated using the CTEQ6L1~\cite{Pumplin:2002vw} parton
distribution functions (PDF) set, except for the \POWHEG \ttbar sample,
for which the CT10 PDF set~\cite{Lai:2010vv} is used. All the samples are then processed further by \PYTHIA,
using the Z2* tune~\cite{Chatrchyan:2011id,Chatrchyan:2013gfi} for simulation of parton showering and subsequent hadronization, and for simulation of the underlying event.
The passage of the particles through the CMS detector
is simulated using the \GEANTfour package~\cite{Agostinelli:2002hh}. All simulated background samples are normalized to the integrated
luminosity of the recorded data, using
inclusive cross sections determined at next-to-leading order, or
next-to-next-to-leading order when available, calculated with
\MCFM v6.6 \cite{MCFM:VJets,MCFM:VV,MCFM:TT,MCFM:SingleTop} and
\FEWZ v3.1 \cite{FEWZ3}, except for the \ttbar sample, for which \textsc{Top++} v2.0~\cite{Czakon:2011xx} is used.

To simulate the signature of interest, we use a model of a generic narrow spin-1 $\PWpr$ resonance
implemented with \MADGRAPH.
We verified that the kinematic distributions agree with those predicted
by implementations of the LH, composite Higgs and HVT models in \MADGRAPH.
The resonance width differs in the three models, but in each case it is found to be negligible with respect to the experimental
resolution. More details on the parameters used for interpretation of the models are given in Section ~\ref{interpretation}.

Extra proton-proton interactions are combined with the generated events before detector simulation to match the observed distribution of the number of additional interactions per bunch crossing (pileup). The simulated samples are also corrected for observed
differences between data and simulation in the efficiencies of the lepton
trigger~\cite{Khachatryan:2014tva}, the lepton identification/isolation~\cite{Khachatryan:2014tva}, and the selection criteria
identifying jets originating from hadronization of bottom quarks (b-tagged jets)~\cite{CMS:BTV13001}.

\section{Reconstruction and selection of events}
\label{sec:RecoAndSel}
\subsection{Trigger and basic event selection}
Candidate events are selected during data taking
using single-lepton triggers, which require either one electron or one muon without
isolation requirements. For electrons the minimum transverse momentum \pt measured at the high level trigger is 80\GeV, while for muons the \pt must be greater than 40\GeV.

After trigger selection, all events are required to have at least one primary-event vertex reconstructed within a 24\unit{cm} window along the beam axis, with a
transverse distance from the nominal pp interaction region
of less than 2\unit{cm}~\cite{CMS:TRK10005}. If more than one identified vertex passes
these requirements, the primary-event vertex is chosen as the one
with the highest sum of $\pt^{2}$ over its constituent tracks.

Individual particle candidates are reconstructed and identified using
the CMS particle-flow (PF) algorithm~\cite{CMS-PAS-PFT-09-001,CMS-PAS-PFT-10-001},
by combining information from all subdetector systems. The reconstructed PF candidates
are each assigned to one of the five candidate
categories: electrons, muons, photons, charged hadrons, and neutral hadrons.

\subsection{Lepton reconstruction and selection}
\label{sec:leptons}

Electron candidates are reconstructed by clustering the energy deposits in the ECAL and then matching the clusters with reconstructed tracks
\cite{Khachatryan:2015hwa}. In order to suppress the multijet background,
electron candidates must pass quality criteria tuned for
high-\pt objects and an isolation selection~\cite{Chatrchyan:2012meb}.
The total scalar sum of the \pt over all
the tracks in a cone of radius $\Delta R = \sqrt{\smash[b]{(\Delta \eta)^2+(\Delta \phi)^2} } = 0.3$ around the electron direction,
excluding tracks within an inner cone of $\Delta R = 0.04$ to remove the contribution
from the electron itself, must be less than 5 \GeV.  A calorimetric isolation parameter is
calculated by summing the energies of reconstructed deposits in both the ECAL and HCAL, not associated with the electron
itself, within a cone of radius
$\Delta R = 0.3$ around the electron. The veto threshold for this
isolation parameter depends on the electron kinematic quantities and the average
amount of additional energy coming from pileup interactions, calculated for each event.
The electron candidates are required to have $\pt > 90$\GeV and $\abs{\eta} < 1.44$ or $1.57<\abs{\eta}<2.5$, thus excluding the transition region between ECAL barrel and endcaps.

{\tolerance=800
Muons are reconstructed with a global fit using both the tracker and muon systems~\cite{CMS:MUO10004}.
An isolation requirement is applied in order to suppress the
background from multijet events in which muons are produced in the semileptonic decay of B hadrons.
A cone of radius $\Delta R = 0.3$ is constructed around
the muon direction. Muon isolation requires that the scalar \pt sum over all tracks
originating from the interaction vertex within the cone, excluding the muon itself, is less than 10\% of the \pt of the muon.
The muon candidates are required to have $\pt > 50$\GeV and $\abs{\eta} < 2.1$
in each selected event.
\par}

Events are required to contain exactly one lepton candidate (electron or muon).
That is, events are rejected if they contain a second lepton candidate with $\pt > 35$\GeV (electrons) or $\pt > 20$\GeV (muons).

\subsection{Jets and missing transverse momentum reconstruction}
\label{sec:jet-met}

Hadronic jets are identified by clustering PF candidates,
using the \textsc{FastJet} v3.0.1 software package~\cite{Cacciari:2011ma}.
In the jet-clustering procedure, charged PF candiates associated with pileup vertices are excluded, to reduce contamination from pileup.
In order to identify a Higgs boson decaying into bottom quarks, jets are clustered using the Cambridge--Aachen algorithm \cite{Wobisch:1998wt} with a
distance parameter of 0.8 (``CA8 jets''). Only the highest \pt CA8 jet is used. Jets in the event are also identified using the anti-\kt jet-clustering algorithm~\cite{Cacciari:2008gp}
with a distance parameter of 0.5 (``AK5 jets''). AK5 jets are required to be separated from the CA8 jet by $\Delta R > 0.8$.
An event-by-event correction based on the projected area of the jet on the front face of
the calorimeter is used to remove the extra energy
deposited in jets by neutral particles coming from pileup.
Furthermore, jet energy corrections are applied, based on measurements
in dijet and photon+jet events in data~\cite{CMS:JetCalibration}.
Additional quality criteria are applied to the jets in order to remove spurious
jet-like features originating from calorimeter noise~\cite{CMS-PAS-JME-10-003}.
The CA8 (AK5) jets are required to be separated from the selected electron or muon candidate by $\Delta R>0.8$ (0.3).
Only jets with $\pt>30$\GeV and
$\abs{\eta}<2.4$ are allowed in the subsequent steps of the analysis.
Furthermore, CA8 jets are not used in the analysis if their pseudorapidity falls in the region $1.0 <\abs{\eta}< 1.8$, thus overlapping the barrel-endcap transition region of the silicon tracker.
In that region, 'noise' can arise when the tracking algorithm reconstructs many fake displaced tracks associated with the jet. The simulation does not sufficiently describe the full material budget of the tracking detector in that region, thus it does not accurately describe this effect.
Without this requirement, a bias can be introduced in the b tagging,
jet substructure and missing transverse momentum information, making this analysis systematically prone to that noise.
The probability of signal events satisfying the requirement that the pseudorapidity of the CA8 jet falls outside the region $1.0 <\abs{\eta}< 1.8$ is 80\% (92\%) for a resonance mass of 1.0 (2.5)\TeV.

A b tagging algorithm, known as the combined secondary vertex
algorithm~\cite{CMS:BTV13001,Chatrchyan:2012jua},
is applied to reconstructed AK5 jets to identify whether they originate from bottom quarks.
This method allows the identification and rejection of the \ttbar events as described in Section~\ref{sec:event-selection}.
The chosen algorithm working point provides a misidentification rate for light-parton jets of $\sim$1\% and an efficiency of $\sim$70\%~\cite{CMS:BTV13001}.
The simulated events are reweighted event-by-event with the ratio of the b tagging efficiency in data and simulation, determined in a sample enriched with b-jets. The average value of the correction factor is 0.95.
The same b tagging algorithm is also used to identify whether the CA8 jet comes from a Higgs boson decaying into bottom quarks, as described in Section~\ref{subsec:Hhadr}.

The missing transverse momentum $\pt^\text{miss}$ is defined as the magnitude of the projection on the plane perpendicular to the beams of the negative vector sum of the momenta of all the reconstructed particles in an event. The raw $\pt^\text{miss}$ value is modified to account for corrections to the energy-momentum scale of all the reconstructed AK5 jets in the event. More details on the $\pt^\text{miss}$ performance in CMS can be found in Refs.~\cite{CMS:METperformances,CMS-PAS-JME-12-002}.
A requirement of $\pt^\text{miss} > 80\,(40)$\GeV is applied for the
electron (muon) channel. The higher threshold for the electron channel
is motivated by the higher contribution from the multijet background expected in the low-$\pt^\text{miss}$ range
due to jets misidentified as electrons.
The background is expected to be negligible in the muon channel, for which a lower $\pt^\text{miss}$ threshold can be used to preserve a higher efficiency for a low-mass signal.

\subsection{The \texorpdfstring{$\PW\to \ell \nu$}{W to l nu} reconstruction and identification}
\label{subsec:Vlept}

The identified electron or muon is associated with the
$\PW \to \ell \nu$ candidate.
The \pt of the undetected neutrino is assumed to be equal
to the $\pt^\text{miss}$.
The longitudinal component $p_{z,\nu}$ of the neutrino momentum is
calculated following a method used originally for the reconstruction
of the invariant mass of the top quark as described in Ref.~\cite{BauerPhd10}.
The method aims to solve a quadratic equation that makes use of the known W boson mass.
Kinematic ambiguities in the solution of the equation are resolved as in Ref.~\cite{BauerPhd10}.
The four-momentum of the neutrino is used to build the four-momentum of the $\PW \to \ell \nu$ candidate.

\subsection{The \texorpdfstring{$\PH \to \bbbar$}{H to b anti-b} identification using jet substructure and b tagging}
\label{subsec:Hhadr}

The CA8 jets are used to reconstruct the jet candidates from
decays of Lorentz-boosted Higgs boson to bottom quarks.
We exploit two techniques to discriminate against quark and gluon jets from the multijet background,
including the requirement that the reconstructed jet mass be close to the Higgs boson mass,
and b tagging methods that discriminate jets originating from the b quarks from those originating from lighter quarks or gluons.

First, we apply a jet-grooming technique~\cite{CMS:SMP12019,JME-13-006} to re-cluster the jet constituents,
while applying additional requirements to remove possible contamination from soft QCD radiation or pileup.
Different jet-grooming algorithms have been explored at CMS, and their performance on jets in
multijet processes has been studied in detail~\cite{CMS:SMP12019}.
In this analysis, we use the \textit{jet pruning} algorithm~\cite{jetpruning1,Ellis:2009me},
which re-clusters each jet starting
from all its original constituents using the CA algorithm iteratively, while discarding soft and large-angle recombinations at each step.
The performance of the algorithm depends on the two parameters, $z_\text{cut}=0.1$ and $D_\text{cut}=\mJ/\ptj$,
which define the maximum allowed hardness and the angle of the recombinations in the clustering algorithm, respectively.
A jet is considered as an H-tagged jet candidate if its pruned mass, \mJ,
computed from the sum of the four-momenta of the constituents
surviving the pruning, falls in the range $110<\mJ<135$\GeV.
The \mJ~window is the result of an optimization based on signal sensitivity and on the constraints due to the higher bounds of the signal regions of
other diboson analyses~\cite{EXO-13-009}.

The simulation modelling of the pruned mass measurement for merged jets from heavy bosons
has been checked using merged $\PW\to\cPaq \Pq'$
decays in $\ttbar$ events with a $\ell$+jets topology~\cite{JME-13-006}.
The data are compared with $\ttbar$ events generated with {\MADGRAPH}, interfaced to \PYTHIA for parton
showering.
The differences between recorded and simulated event samples in the pruned jet mass scale
and resolution are found to be up to 1.7\% and 11\%, respectively.
In addition, the modelling of bottom quark fragmentation is checked through reconstruction of the
top quark mass in these $\ttbar$ events~\cite{JME-13-007}.

To discriminate between quark and gluon jets, on one hand, and a Higgs-initiated jet, on the other, formed by the hadronization of two bottom quarks, we use a H tagging technique~\cite{CMS:BTV13001}. This procedure splits the candidate H-jet into two sub-jets by reversing the
last step of the CA8 pruning recombination algorithm.
Depending on the angular separation $\Delta R$ of the two sub-jets, different
b tagging discriminators are used to tag the H-jet candidate. If $\Delta R>0.3$, then the b tagging algorithm is applied to both of the individual sub-jets of the CA8 jet; otherwise,
it is applied to the whole CA8 jet.
The chosen algorithm working point provides a misidentification rate of 10\% and an efficiency of 80\%.
The ratio of the b tagging efficiency between data and simulation, in a sample enriched with b-jets
from gluon splitting by requiring two muons within the CA8 jet, is used to reweight the simulated events.

\subsection{Final event selection and categorization}
\label{sec:event-selection}

After reconstructing the W and Higgs bosons, we apply the final
selections used
for the search. Both the W and Higgs boson
candidates must have a \pt greater than 200\GeV.
In addition, we apply topological selection criteria,
requiring that the W and Higgs bosons are approximately back-to-back,
since they tend to be isotropically distributed for background events.
In particular, the $\Delta R$ distance between the lepton and the H-tagged jet must be greater than $\pi/2$, the azimuthal angular separation between the $\pt^\text{miss}$ and the H-tagged jet must be greater than 2.0 radians, and the azimuthal angular separation between the $\PW \to \ell \nu$
and H-tagged jet candidates must be greater than 2.0 radians.
To further reduce the level of the \ttbar background,
events with one or more reconstructed AK5 jets, not overlapping
with the CA8 H-tagged jet candidate as described previously in Section~\ref{sec:jet-met}, are analyzed.
If one or more of the AK5 jets is b-tagged, the event is rejected.
Furthermore, a leptonically decaying top quark candidate mass $m_\text{top}^\ell$ is reconstructed from the lepton, $\pt^\text{miss}$, and the closest AK5 jet to the lepton
using the method described in Ref.~\cite{BauerPhd10}.
A hadronically decaying top quark candidate mass $m_\text{top}^\mathrm{h}$ is reconstructed from the CA8 H-tagged jet candidate and the closest AK5 jet.
Events with $120<m_\text{top}^\ell<240$\GeV or $160<m_\text{top}^\mathrm{h}<280$\GeV are rejected.
The chosen windows around the top quark mass are the result of an optimization carried out in this analysis,
taking into account the asymmetric tails at larger values due to combinatorial background.
If several distinct WH resonance candidates are present in the same event,
only the candidate with the highest-\pt H-tagged jet is kept for further analysis.
The invariant mass of the WH resonance (\mWH) is required to be at least 0.7\TeV.
The signal efficiency for the full event selection ranges between $\sim$3\% and $\sim$9\%, depending on the resonance mass.

\section{Modelling of background and signal}

\subsection{Background estimation}
\label{sec:bkgd}

After the full event selection, the two dominant remaining backgrounds are expected to come from W+jets and \ttbar events.
Backgrounds from \ttbar, single top quark, and diboson production are estimated using simulated samples
after applying correction factors derived from control samples in data.
For the W+jets background estimation, a procedure based on data has been developed
to determine both the normalization and the \mWH shape.

\begin{figure}[htbp]
\centering
\includegraphics[width=0.49\textwidth]{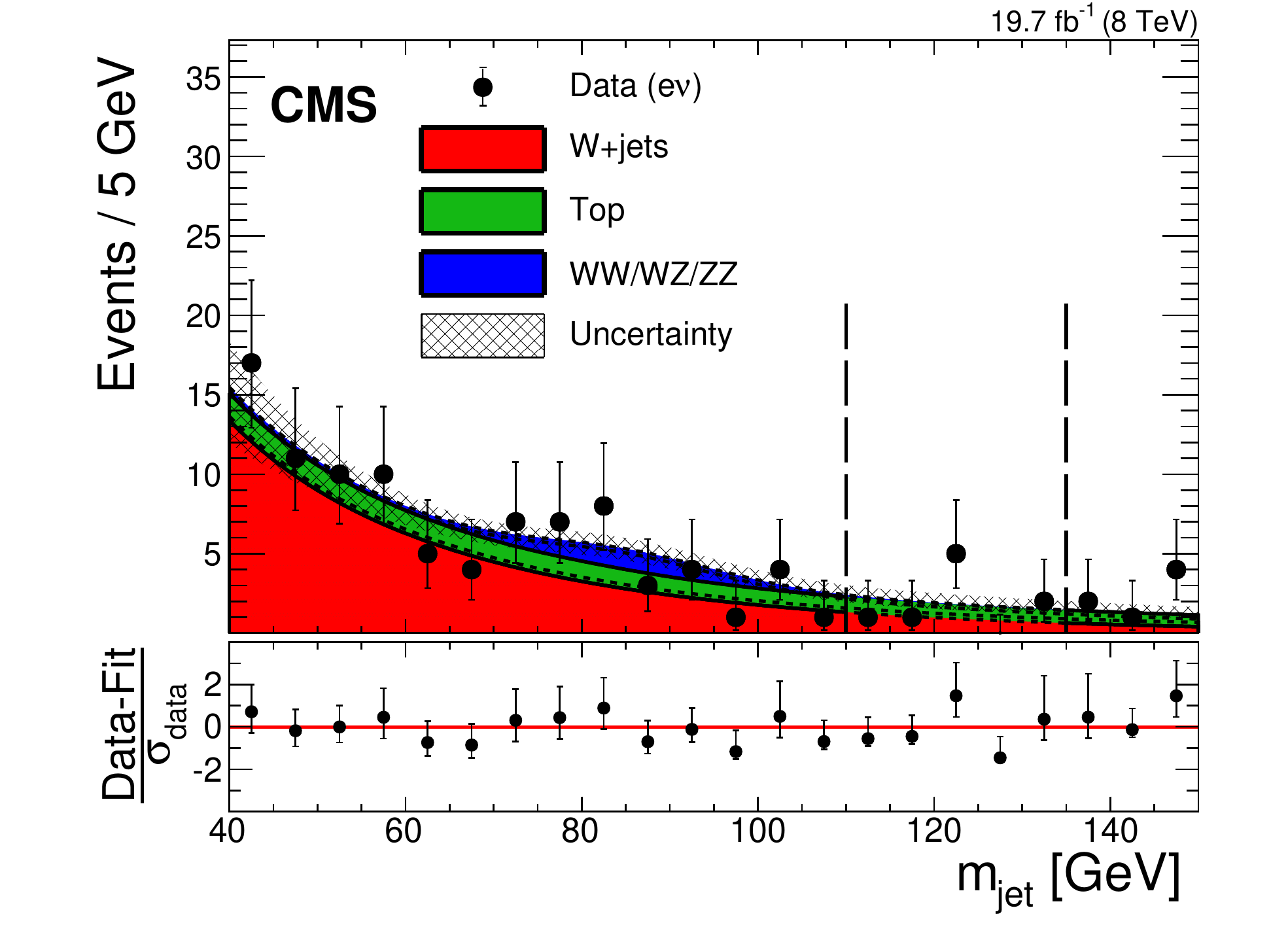}
\includegraphics[width=0.49\textwidth]{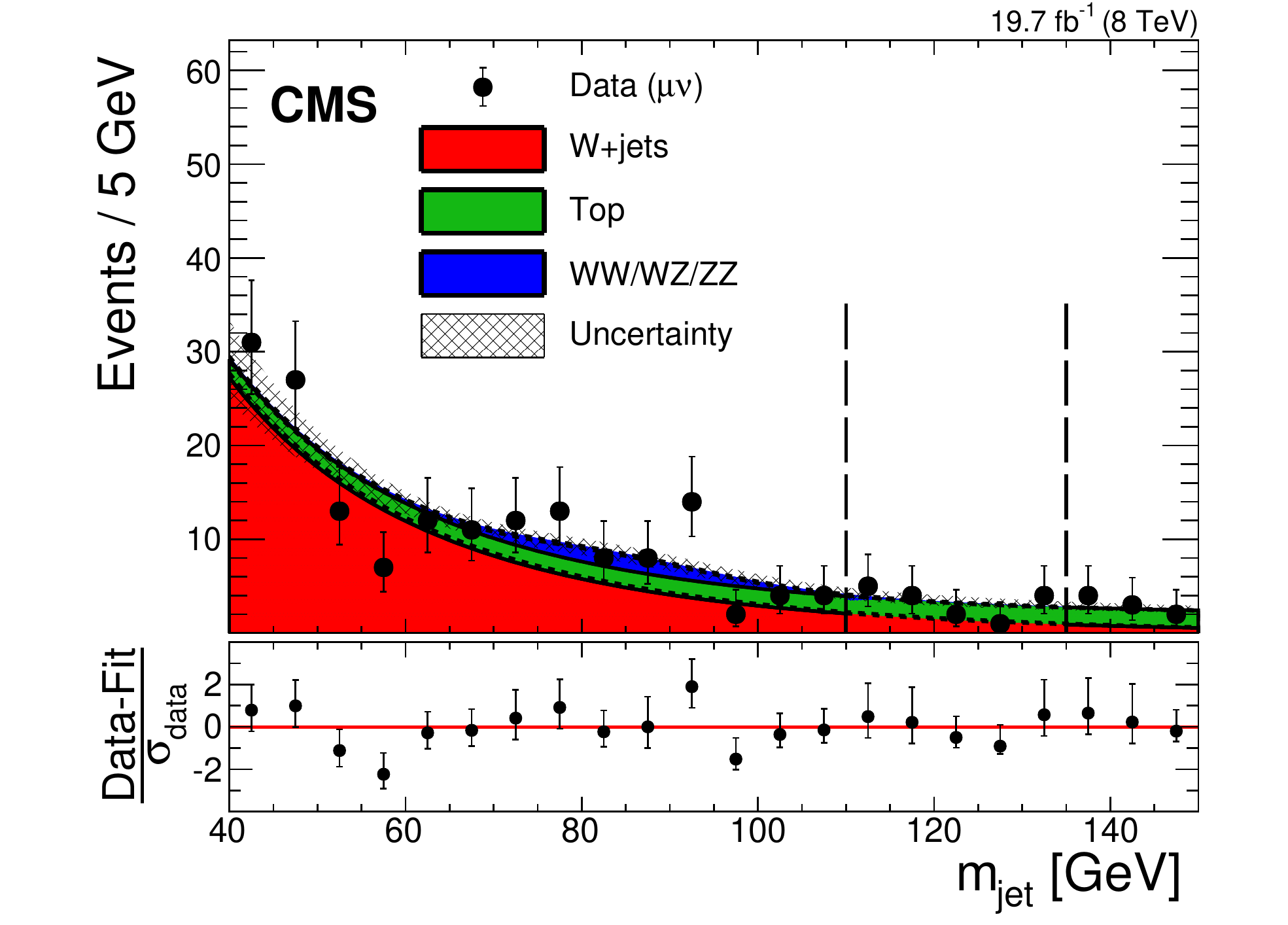}
\caption{Distributions of the pruned jet mass, \mJ, in the electron (\cmsLeft) and muon (\cmsRight) channels. The signal region lies between the dashed vertical lines. The hatched region indicates the statistical uncertainty of the fit. At the bottom of each plot, the bin-by-bin fit residuals, $(\text{Data}-\text{Fit})/\sigma_\text{data}$,
are shown. }
\label{fig:WJetsNormalization}
\end{figure}

For the W+jets normalization estimate, a signal-depleted control region is defined outside the
\mJ~mass window described in Section~\ref{subsec:Hhadr}.
A lower sideband region is defined in the \mJ~range [40, 110]\GeV as well as an upper
sideband in the range [135, 150]\GeV.
The overall normalization of the W+jets background in the signal
region is determined from the likelihood of the sum of backgrounds fit to the \mJ~distribution
in both sidebands of the observed data. In this approach, simulated events are
replaced by an analytical function, which has been determined individually for each
background process.
Figure~\ref{fig:WJetsNormalization} shows the result of this fit procedure, where all selections are applied
except the final \mJ~signal window requirement. The inclusive W+jets
background is predicted from a fit excluding the signal region (between the vertical dashed lines),
while the other backgrounds are estimated from simulation.

The shape of the W+jets background as a function of \mWH in the signal region is estimated using the lower sideband region of the \mJ~distribution.
Correlations needed to extrapolate from the sideband to the signal region are determined from simulation
through an extrapolation function defined as:

\begin{equation}
\alpha_\mathrm{MC}(\mWH) = \frac{F_\mathrm{MC, SR}^{\PW+\text{jets}}(\mWH)}{F_\mathrm{MC, SB}^{\PW+\text{jets}}(\mWH)},
\end{equation}
where $F_\mathrm{MC, SR}^{\PW+\text{jets}}$ and $F_\mathrm{MC,
SB}^{\PW+\text{jets}}$ are the probability density functions determined from
the \mWH spectrum in simulation for the signal region and
low-\mJ~sideband region, respectively.

In order to estimate the W+jets contribution $F_{\text{DATA}, \mathrm{SB}}^{\text{W+jets}}$ in the control region of the
data the other backgrounds
are subtracted from the observed \mWH distribution in the lower sideband region.
The shape of the W+jets
background distribution in the signal region is obtained by scaling $F_{\text{DATA},
\mathrm{SB}}^{\PW+\text{jets}}$ according to $\alpha_\mathrm{MC}$. The final
prediction of the background contribution in the signal region,
$N^\text{BKGD}_\mathrm{SR}$, is given by
\ifthenelse{\boolean{cms@external}}{
\begin{multline}
N^\mathrm{BKGD}_\mathrm{SR}(\mWH) = C_{\mathrm{SR}}^{\PW+\text{jets}}\, F_{\text{DATA}, \mathrm{SB}}^{\PW+\text{jets}}(\mWH)\,
\alpha_\mathrm{MC}(\mWH)\\
 + \sum_{k} C_{\mathrm{SR}}^{k}~F_\mathrm{MC, SR}^{k}(\mWH),
\end{multline}
}{
\begin{equation}
N^\mathrm{BKGD}_\mathrm{SR}(\mWH) = C_{\mathrm{SR}}^{\PW+\text{jets}}\, F_{\text{DATA}, \mathrm{SB}}^{\PW+\text{jets}}(\mWH)\,
\alpha_\mathrm{MC}(\mWH) + \sum_{k} C_{\mathrm{SR}}^{k}~F_\mathrm{MC, SR}^{k}(\mWH),
\end{equation}
}
where the index $k$ runs over the list of minor backgrounds, and
$C_{\mathrm{SR}}^{\PW+\text{jets}}$ and $C_{\mathrm{SR}}^{k}$
represent the normalizations of the yields of the dominant W+jets background and of the
different minor background contributions. The $C_{\mathrm{SR}}^{\PW+\text{jets}}$ parameter is determined from the
fit to the \mJ~distribution as described above, while each $C_{\mathrm{SR}}^{k}$ is determined from simulation.
The ratio $\alpha_\mathrm{MC}$ accounts for the small kinematic differences between signal and sideband regions,
and is largely independent of the assumptions on the overall cross section.
The validity and robustness of this method have been studied in data using a
lower \mJ\ sideband of [40, 80]\GeV to predict an alternate signal region
with \mJ\ in the range [80, 110]\GeV. Both the normalization and shape of the W+jets background are successfully estimated for the alternate signal region.
This alternate signal region differs from the signal region of the
search for WW or WZ resonances in Ref.~\cite{EXO-13-009} as b tagging is applied to the CA8 jet.
We are therefore able to evaluate the potential WW and WZ signal contamination
in the alternate signal region and find less than 5\% signal contamination, assuming a signal cross section corresponding to the exclusion limit for a WW resonance from Ref.~\cite{EXO-13-009}.
The \mWH distribution of the background in the signal and lower sideband regions is described analytically by a
function defined as $f(x)\propto \exp[-x/(c_0+c_{1}x)]$, which is found to describe the simulation well.
Alternative fit functions have been studied but in all cases the background shapes agree with that of the default function within uncertainties.

\begin{figure}[htbp]
\centering
\includegraphics[width=0.49\textwidth]{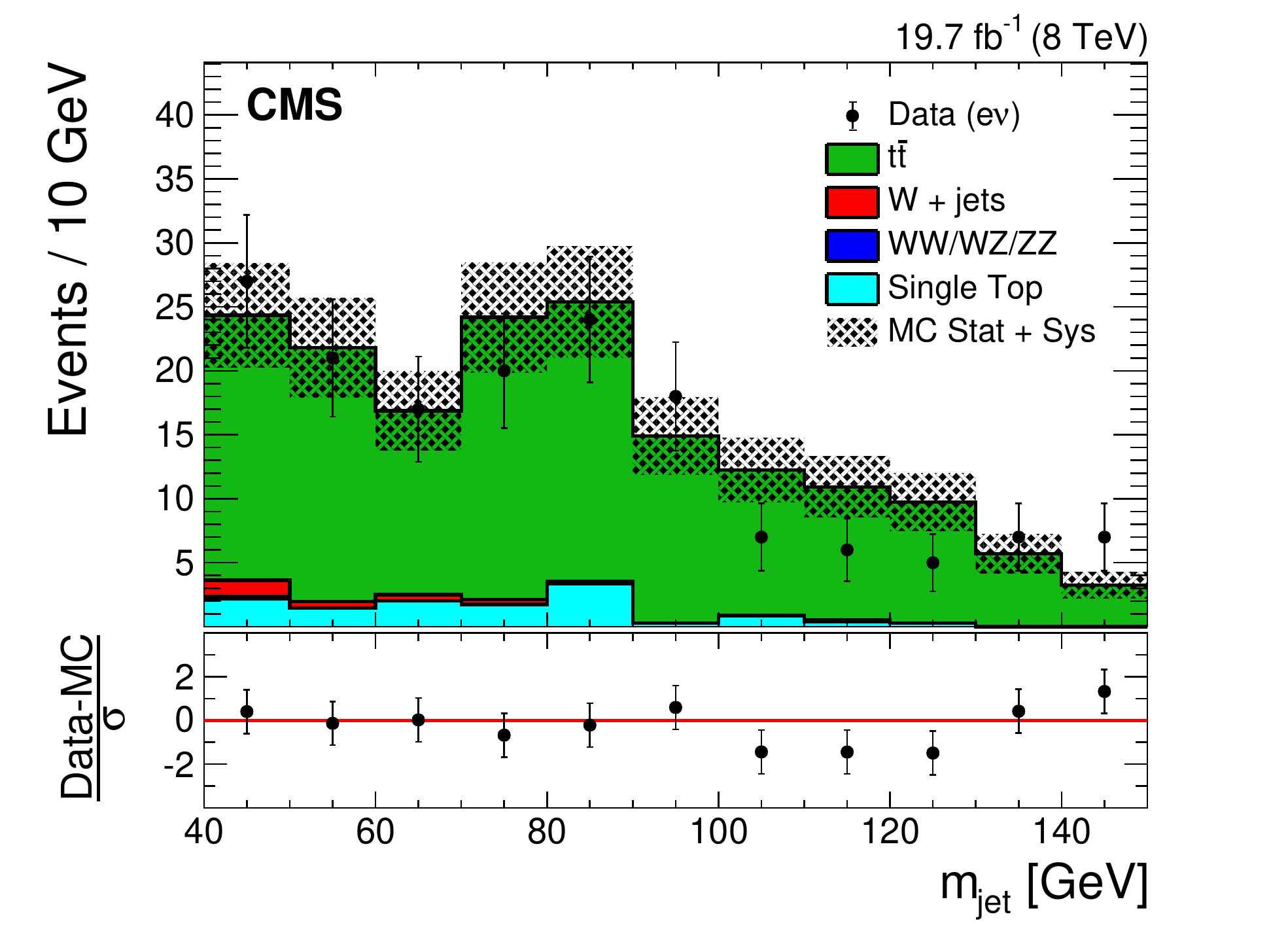}
\includegraphics[width=0.49\textwidth]{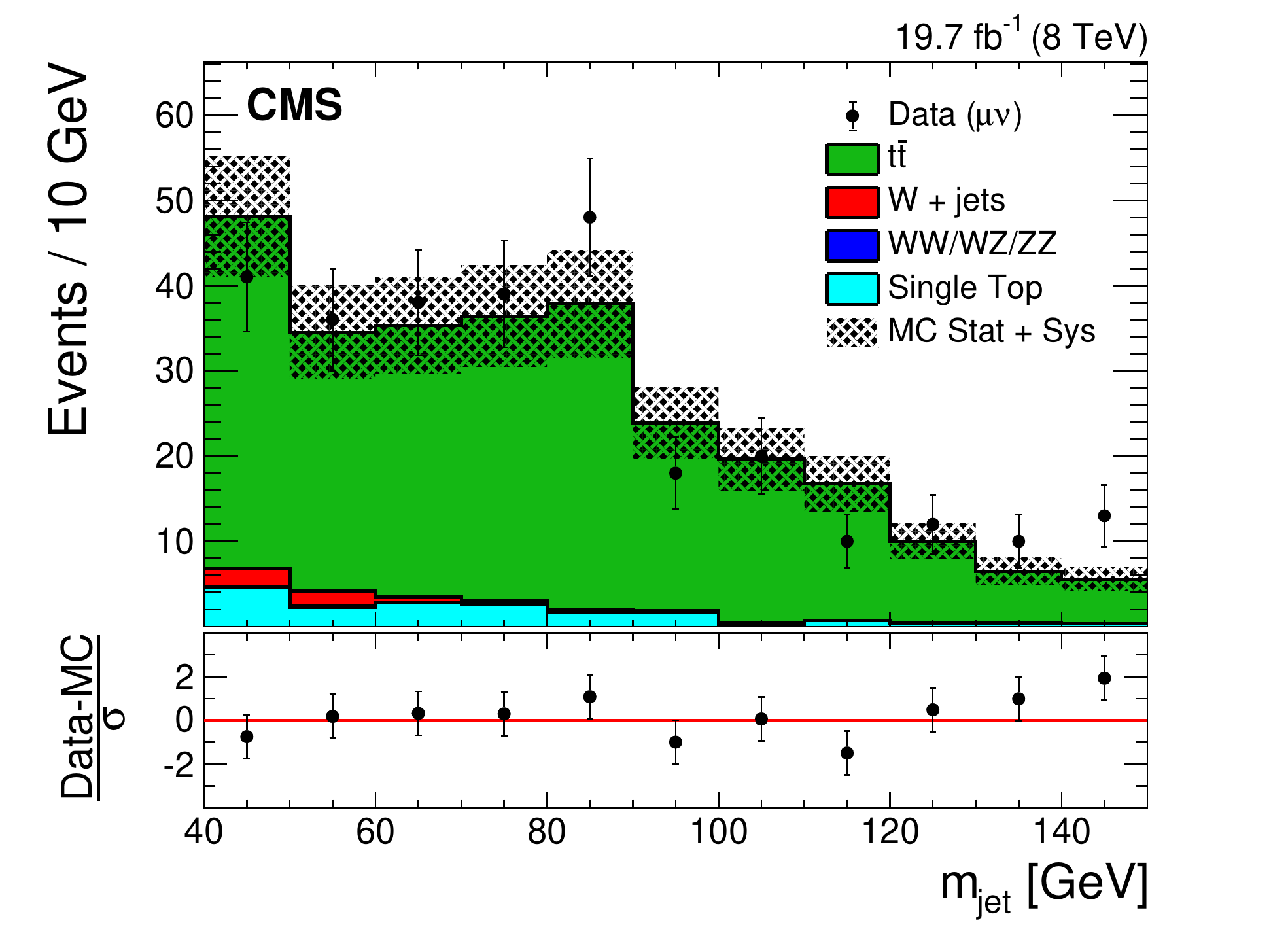}
\caption{Distributions of \mJ~in the top quark enriched control sample in the
electron (\cmsLeft) and muon (\cmsRight) channels. The hatched region indicates the overall uncertainty in the background. In the lower panels, the bin-by-bin residuals,
$(\text{Data}-\mathrm{MC})/\sigma$ are shown, where $\sigma$ is the sum in quadrature of the statistical uncertainty
of the data, the simulation, and the systematic uncertainty in the \ttbar background.}
\label{fig:ttbarControlCut}
\end{figure}

For the \ttbar background estimate, a control sample is selected
by applying all analysis requirements, except that the b-tagged jet veto is inverted, the veto on the top quark mass is dropped,
and the \mJ~requirement is removed.
The data are compared with the predictions from simulation and good agreement is found.
The pruned jet mass distribution in the top quark enriched control sample is
shown in Fig.~\ref{fig:ttbarControlCut}.
The pruned jet mass distribution shows a small peak due to isolated W boson decays into hadrons,
along with a smoothly varying combinatorial component mainly due to events in which the extra b-tagged jet
from the top quark decay is in the proximity of the W boson.
The difference in normalization between data and simulation is found to be $4.6\pm5.6$\%, where the quoted uncertainty is only statistical. This normalization difference is applied to correct the normalization of \ttbar background
in the signal region. The relative uncertainty of 5.6\% is used to quantify the uncertainty in the
\ttbar and single top quark background normalization, as described in Section~\ref{sec:syst:background}.

\subsection{Modelling of the signal mass distribution}
\label{sec:signal}

The shape of the reconstructed signal mass distribution is extracted from the simulated signal samples.
In the final analysis of the
\mWH spectrum, the statistical signal sensitivity depends
on an accurate description of the signal shape.
The signal shape is parametrized with a double-sided Crystal Ball function (\ie\ a Gaussian core with power-law tails on both
sides)~\cite{CrystalBallRef} to describe the CMS detector resolution. Figure 4 shows an example of this parametrization for a $\PWpr$ mass of 1.5\TeV.
To take into account differences between the electron and muon \pt resolutions at high \pt,
the signal mass distribution is parametrized separately for events with electrons and muons.
The resolution of the reconstructed \mWH is given by the width of the Gaussian core and is found to be 4--6\%.

\section{Systematic uncertainties}

\subsection{Systematic uncertainties in the background estimation}
\label{sec:syst:background}
Uncertainties in the estimation of the background affect both the
normalization and the shape of the \mWH distribution.
The systematic uncertainty in the W+jets background yield is
dominated by the statistical uncertainty associated with the number of
events in data in the $m_{\text{jet}}$ sideband regions, and it is found to be about 59\%\,(42\%) in the electron (muon) channel. The systematic uncertainty in the
\ttbar~normalization comes from the data-to-simulation ratio derived in the top-quark-enriched control
sample (\TTBARNORMUNCERT) as described in Section~\ref{sec:bkgd}.
The systematic uncertainties in the WW, \PW\Zo, and ZZ inclusive cross sections are assigned to be
\VVNORMUNCERT, taken from the relative difference in the mean value
between the CMS WW cross section measurement at $\sqrt{s}=8$\TeV
and the SM expectation~\cite{Chatrchyan:2013oev}.

{\tolerance=800
Systematic uncertainties in the W+jets background shape are estimated from the
covariance matrix of the fit to the extrapolated data sideband and from the
uncertainties in the modelling of $\alpha_\mathrm{MC}(\mWH)$.
They are driven by the available data in the
sidebands and the number of events generated for the simulation of the
W+jets background, respectively. These uncertainties are shown in Fig.~\ref{fig:MZZwithBackgroundWH}, and they are found to be about 30\% (120\%) at $\mWH \approx 1\TeV$\,(1.8\TeV). The estimation of the systematic uncertainty in the shape of the \ttbar background takes into account the following contributions: the statistical uncertainty associated with the simulated event sample, the choices of regularization/factorization scales (varied up and down by a factor of 2), the matching scales in the \MADGRAPH simulation, and an observed difference between \MADGRAPH and \POWHEG simulations.
\par}

Systematic effects from rare noise events identified in the tracker overlap
region were specifically studied in the context of the
acceptance requirement introduced for \PH-jet candidates ($\abs{\eta} < 1.0$ or $\abs{\eta} > 1.8$)
as described in Section~\ref{sec:RecoAndSel}. Those studies conclude that any residual noise
effects following the imposition of this requirement are negligible.
No additional source of systematic uncertainty is taken into account for the background predictions.

\subsection{Systematic uncertainties in the signal prediction}

Systematic uncertainties in the signal prediction affect both the signal efficiency and the
\mWH shape. The primary uncertainties in signal yields are summarized in Table~\ref{table:systematics} and described below.

The systematic uncertainties in the signal efficiency due to the electron energy (E) and muon \pt scales are evaluated by varying the lepton E or \pt within one standard deviation
of the corresponding uncertainty~\cite{CMS:MUO10004,Khachatryan:2015hwa};
the uncertainties due to the electron E and muon \pt resolutions
are estimated applying a \pt and E smearing, respectively.
In this process, variations in the lepton E or \pt are propagated consistently to the $\pt^\text{miss}$ vector.
We also take into account the systematic uncertainties affecting the observed-to-simulated scale factors for the efficiencies of the lepton trigger,
identification and isolation requirements. These efficiencies are derived using a specialized
tag-and-probe analysis with $\cPZ\to \ell^{+}\ell^{-}$ events~\cite{Khachatryan:2010xn}, and the uncertainty in the ratio of the efficiencies is
taken as the systematic uncertainty. The uncertainties in the efficiencies of the electron (muon) trigger and the electron (muon) identification with isolation are 3\%\,(3\%) and 3\%\,(4\%), respectively.

The signal efficiency is also affected by the uncertainties in the jet energy-momentum scale and resolution.
The jet energy-momentum scale and resolution are varied within their \pt- and $\eta$-dependent
uncertainties~\cite{CMS:JetCalibration} to estimate their impact on the signal efficiency. The variations are also propagated consistently to the $\pt^\text{miss}$ vector.

The momentum scale uncertainty of particles that are not identified as leptons or clustered in jets (`unclustered energy-momentum') is found to introduce an uncertainty of less than 0.5\% in the signal efficiency.

We also include systematic uncertainties in the signal efficiency due to
uncertainties in data-to-simulation scale factors for the pruned jet mass
tagging, derived from the top quark enriched control sample~\cite{JME-13-006}
and b-tagged jet identification efficiencies~\cite{CMS:BTV13001}.
These sources introduce a systematic uncertainty in the mass tagging and b tagging of the Higgs boson
of 2--10\% and 2--8\%, respectively, depending on the signal mass.

The systematic uncertainty due to the modelling of pileup is estimated by reweighting
the signal simulation samples such that the distribution of the number of interactions
per bunch crossing is shifted according to the uncertainty in the
inelastic proton-proton cross section~\cite{Chatrchyan:2012nj,TOTEM:2013}.

The impact of the proton PDF uncertainties on the signal efficiency is evaluated
with the PDF4LHC prescription~\cite{Botje:2011sn,Alekhin:2011sk}, using
the MSTW2008~\cite{MSTW}
and NNPDF2.1~\cite{NNPDF} PDF sets.
The uncertainty in the integrated luminosity is 2.6\% \cite{CMS:LUM13001}.

\begin{table*}[htb]
\centering
\topcaption{Summary of the systematic uncertainties in the signal yield, relative to the expected number of events.
}
\label{table:systematics}
\small
\begin{tabular}{lccc}
\multirow{2}{*}{{Source}}     &  \multicolumn{2}{c}{{Uncertainty [\%]}}\\
                                     & electron     & muon \\
\hline
\hline
Lepton trigger and ID efficiencies     & {3}      & {2}      \T\\
Lepton \pt scale               & {$<$0.5} & {1}       \T\\
Lepton \pt resolution          & {$<$0.1} & {$<$0.1}       \T\\
Jet energy-momentum scale                  & \multicolumn{2}{c}{1--3}\T\\
Jet energy-momentum resolution             & \multicolumn{2}{c}{$<$0.5}\T\\
Higgs boson mass tagging efficiency        & \multicolumn{2}{c}{2--10}\T\\
Higgs boson b tagging efficiency           & \multicolumn{2}{c}{2--8}\T\\
Unclustered energy scale                   & \multicolumn{2}{c}{$<$0.5}\T\\
Pileup                                     & \multicolumn{2}{c}{0.5}\T\\
PDF                                        & \multicolumn{2}{c}{$<$0.5}\T\\
Integrated luminosity                      & \multicolumn{2}{c}{\LUMIUNCERT}\T\\
\hline
\end{tabular}
\end{table*}

In addition to systematic uncertainties in the signal efficiency discussed above,
we consider uncertainties in the signal resonance peak position and width.
The systematic effects that could change the signal shape are
the uncertainties due to the \pt/energy-momentum scale and resolution of electrons, muons, jets,
and the unclustered energy-momentum scale. For each of these sources of
experimental uncertainty, the energy-momentum of the lepton and jets, as well as the corresponding $\pt^\text{miss}$
vector, are varied (or smeared) by their relative uncertainties.
The uncertainty in the peak position of the signal is estimated
to be less than 1\%.
The jet energy-momentum scale and
resolution introduce a relative uncertainty of about 3\% in the
signal width.
The unclustered
energy-momentum scale introduces an uncertainty in the signal width of 1\% at lower resonance masses ($<$1.5\TeV),
and of 3\% at higher masses.

\section{Results}
\label{sec:results}

The predicted number of background events in the signal
region after the inclusion of all backgrounds is summarized in Table~\ref{table:WHExpectedYields}
and compared with observations.
The yields are quoted in the range $0.7 < \mWH <  3$\TeV.
The expected background is derived with the
sideband procedure.  The uncertainties in the background prediction from data are statistical in nature, as they depend on
the number of events in the sideband region.
The muon channel has more expected background events than the electron channel owing to the lower $\pt^\text{miss}$ requirement on the muon and its worse mass resolution at high $\pt$.

\begin{figure}[htbp]
\centering
\includegraphics[width=0.49\textwidth]{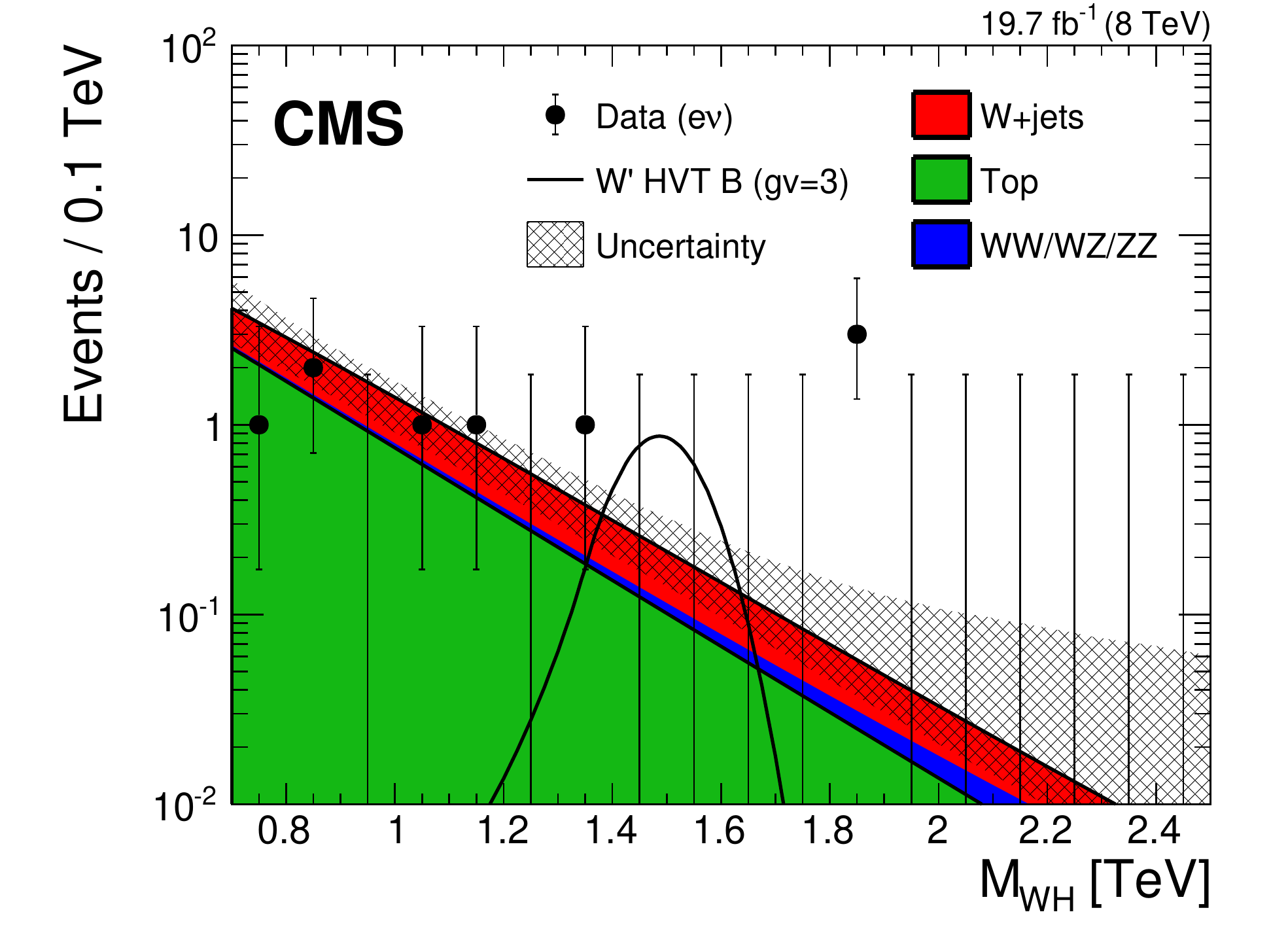}
\includegraphics[width=0.49\textwidth]{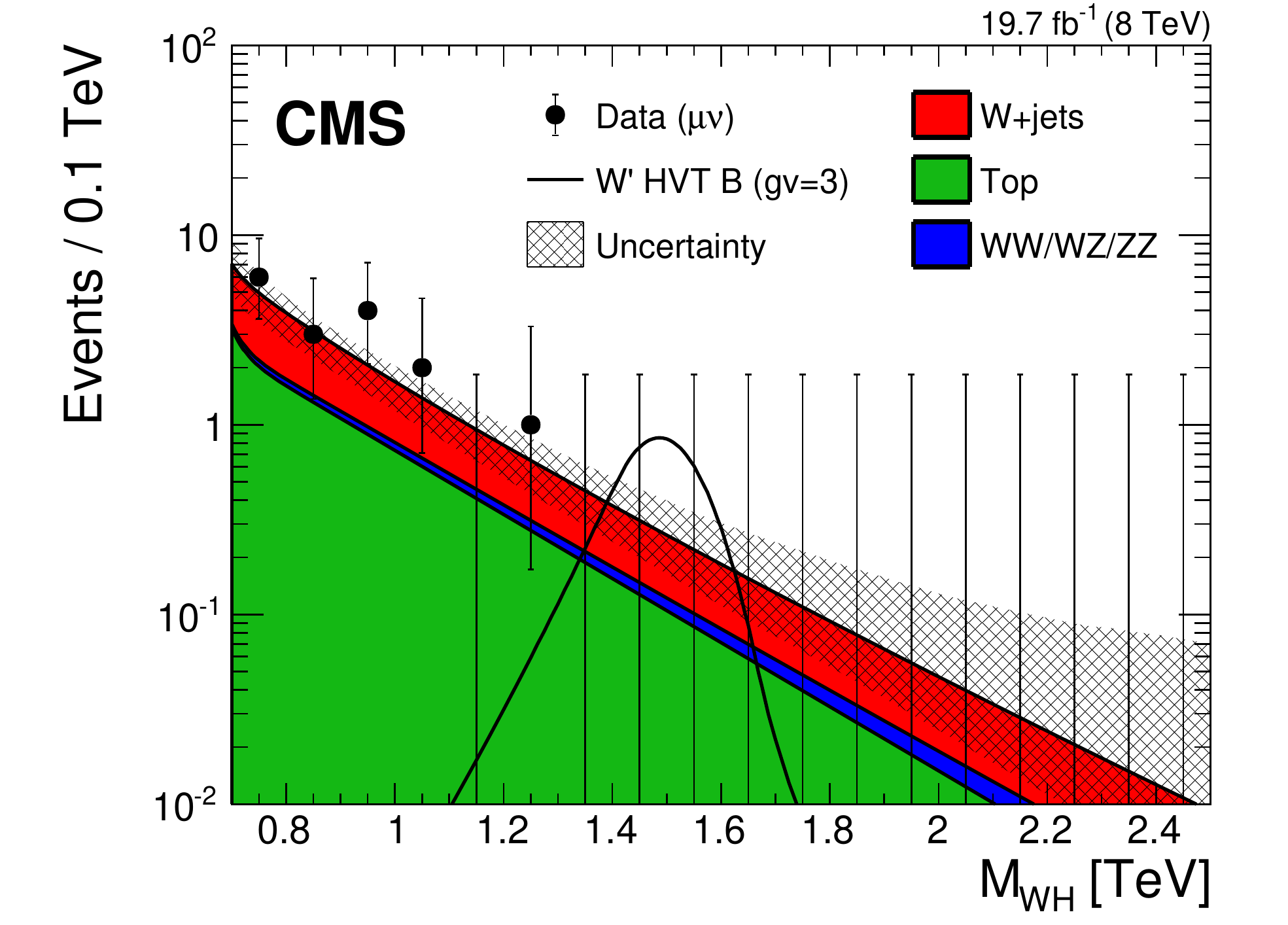}
\caption{
Final distributions in \mWH for data and expected backgrounds for electron (\cmsLeft) and muon (\cmsRight) categories.
The 68\% error bars for Poisson event counts are obtained from the Neyman construction~\cite{Garwood}. The hatched region indicates the statistical uncertainty of the fit combined with the systematical uncertainty in the shape. This figure also shows a hypothetical $\PWpr$ signal with mass of 1.5\TeV, normalized to the cross section predicted by the HVT model B with parameter $g_\Vo=3$ as described in Section~\ref{cross_section_limits}.
}
\label{fig:MZZwithBackgroundWH}
\end{figure}

Figure~\ref{fig:MZZwithBackgroundWH} shows the \mWH spectra after all selection criteria have been applied.
The highest mass event is in the electron category and has $\mWH\approx 1.9$\TeV.
The observed data and the predicted background in the muon channel agree.
In the electron channel, an excess of three events is observed with $\mWH > 1.8$\TeV,
where about 0.3 events are expected, while in the muon channel no events with $\mWH > 1.8$\TeV are observed, where about 0.3 events are expected.

\begin{table}[htbp]
\centering
\topcaption{
Observed and expected yields in the signal region together with statistical uncertainties.
}
\label{table:WHExpectedYields}
\begin{tabular}{lcc}
   & $\Pe\nu$+H-jet & {$\mu\nu$+H-jet}   \\
\hline \hline
 Observed yield     & 9   & 16  \\
 Expected total background   & $11.3 \pm 3.1$  & $14.9 \pm 3.1$   \\
\hline
 W+jets   & $4.7 \pm 2.9$  & $7.0 \pm 3.1$   \\
 Top  & $6.3 \pm 1.1$ & $7.3 \pm 0.4$ \\
 VV   & $0.4 \pm 0.1$  & $0.6 \pm 0.2$   \\
\hline \hline
\end{tabular}
\end{table}

\section{Statistical and model interpretation}
\label{interpretation}

\subsection{Significance of the data}

A comparison between the \mWH\ distribution observed in data and the largely data-driven background prediction is used to test for the presence of a
resonance decaying into WH.
The statistical test is performed based on a profile likelihood discriminant
that describes an unbinned shape analysis.
Systematic uncertainties in the signal and background yields
are treated as nuisance parameters and profiled in the statistical
interpretation using log-normal priors.

\begin{figure}[htb]
\centering
     \includegraphics[width=0.49\textwidth]{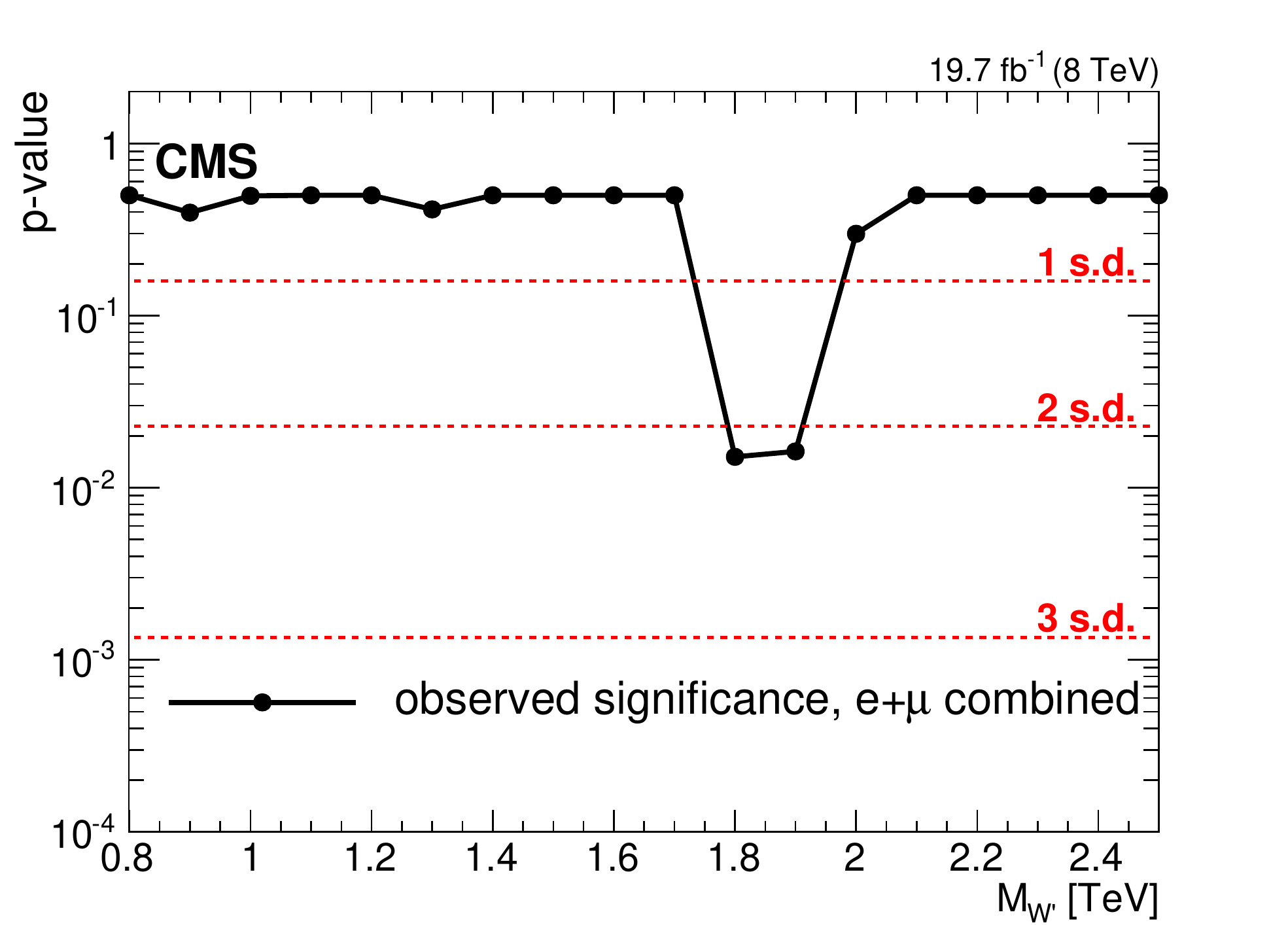}
\caption{
  Local p-value of the combined electron and muon data as a function of the $\PWpr$ boson mass,
  probing a narrow $\PW\PH$ resonance.
}
\label{fig:sig}
\end{figure}

We evaluate the local significance of the observations in the context of the described test, under the assumptions of
a narrow resonance decaying into the WH final state and lepton universality for the W boson decay, by combining the two event categories. Correlations arising from the uncertainties common to both channels are taken into account. The result is shown in Fig.~\ref{fig:sig}.
The highest local significance of 2.2 standard deviations is found for a resonance mass of 1.8\TeV,
driven by the excess in the electron channel described in Section~\ref{sec:results}. The corresponding local significance for a resonance of 1.8\TeV in the electron channel is 2.9 standard deviations, while in the muon channel there is no significance. Taking into account the look-elsewhere effect~\cite{lookelse}, a local significance of 2.9 standard deviations translates into a global significance of about 1.9 standard deviations searching for resonances over the full mass range 0.8--2.5\TeV and across two channels. We conclude that the results are thus statistically compatible with the SM expectation within 2
standard deviations.

\subsection{Cross section limits}
\label{cross_section_limits}

We set upper limits on the production cross section of a new resonance following
the modified-frequentist \CLs method~\cite{CLs1,Junk:1999kv}.
Exclusion limits can be set as a function of the $\PWpr$ boson mass,
under the narrow-width approximation.
The results are interpreted in the HVT model B~\cite{Pappadopulo:2014qza} which mimics the properties
of composite Higgs scenarios, and in the context of the little Higgs model~\cite{Han:2003wu}.
Typical parameter values for the HVT model B are
\begin{equation}\label{eqn:HVTpars}
\abs{c_\PH} \approx \abs{c_\mathrm{F}}\approx 1, \quad  g_\Vo \geq 3,
\end{equation}
where $c_\PH$ describes interactions involving the Higgs boson or longitudinally polarized SM vector bosons,
$c_\mathrm{F}$ describes the direct interactions of the $\PWpr$ with fermions, and $g_\Vo$ is the typical strength of the new interaction.
In this scenario, decays of the $\PWpr$ boson into a diboson are dominant and the $\PWpr \to$ WH branching fraction is almost equal
to that of the decay into WZ. The parameter points for this scenario are currently not well constrained
from experiments~\cite{Pappadopulo:2014qza} because of the suppressed fermionic couplings of the $\PWpr$ boson.

The following parameters are used for interpretation of the results:
$g_\Vo = 3$, $c_\PH = -1$ and $c_\mathrm{F} = 1$ in the HVT model B and $\cot{2\theta} =  2.3$,
$\cot{\theta} = -0.20799$ in the LH model, where $\theta$ is a mixing angle parameter
that determines $\PWpr$ couplings and that $\cot{2\theta}$ and $\cot{\theta}$ can be directly
related to $c_\PH$ and $c_\mathrm{F}$.

The intrinsic width and cross section for both models are listed in Table~\ref{tab:gamma_xsec_LH_HVT} for
several resonance masses.
The widths for the HVT model B are computed by means of Eqs. (2.25) and (2.31) in Ref.~\cite{Pappadopulo:2014qza},
while the cross sections were obtained using the online tools provided by the authors of Ref.~\cite{Pappadopulo:2014qza}.
The width is less than 5\% for the following parameter values: $0.95<g_\Vo<3.76$, $c_\PH = -1$, and $c_\mathrm{F} = 1$; $g_\Vo<3.9$, $c_\PH = -1$, and $c_\mathrm{F} = 0$; or
$g_\Vo<7.8$, $c_\PH = 0.5$, and $c_\mathrm{F} = 0$. The widths for the LH model have been computed by means of Eq. (15) in Ref.~\cite{Burdman:2002ns},
and they are less than 5\% for values of $0.084<\abs{\cot{\theta}}<1.21$. Hence, in both models we can consider the width to be negligible compared to the experimental resolution.

\begin{table*}[htb]
\topcaption{
Intrinsic total widths ($\Gamma$) and cross sections ($\sigma$) for the LH model and HVT model B for different resonance masses.
The $\PW\PH\to \ell \nu {\bbbar}$ branching fraction is not included in the calculation.
}
\label{tab:gamma_xsec_LH_HVT}
\begin{center}
\small
\begin{tabular}{ccccc}
\multirow{2}{*}{{Resonance mass [\TeVns{}]}} & \multicolumn{2}{c}{{LH model}} & \multicolumn{2}{c}{{HVT model B}} \\
                 & $\Gamma$ [\GeVns{}] & $\sigma$ [pb] & $\Gamma$ [\GeVns{}] & $\sigma$ [pb] \\
\hline \hline
0.8 & 7.22 & 5.09$\times 10^{-1}$ & 24.1 & 3.37$\times 10^{-1}$ \\
0.9 & 8.12 & 3.03$\times 10^{-1}$ & 27.1 & 2.48$\times 10^{-1}$ \\
1.0 & 9.02 & 1.87$\times 10^{-1}$ & 30.1 & 1.71$\times 10^{-1}$ \\
1.1 & 9.92 & 1.18$\times 10^{-1}$ & 33.1 & 1.16$\times 10^{-1}$ \\
1.2 & 10.8 & 7.65$\times 10^{-2}$ & 36.1 & 8.05$\times 10^{-2}$ \\
1.3 & 11.7 & 5.06$\times 10^{-2}$ & 39.1 & 5.59$\times 10^{-2}$ \\
1.4 & 12.6 & 3.39$\times 10^{-2}$ & 42.2 & 3.88$\times 10^{-2}$ \\
1.5 & 13.5 & 2.29$\times 10^{-2}$ & 45.2 & 2.51$\times 10^{-2}$ \\
1.6 & 14.4 & 1.56$\times 10^{-2}$ & 48.2 & 1.87$\times 10^{-2}$ \\
1.7 & 15.3 & 1.08$\times 10^{-2}$ & 51.2 & 1.30$\times 10^{-2}$ \\
1.8 & 16.2 & 7.43$\times 10^{-3}$ & 54.2 & 9.03$\times 10^{-3}$ \\
1.9 & 17.1 & 5.17$\times 10^{-3}$ & 57.2 & 6.27$\times 10^{-3}$ \\
2.0 & 18.0 & 3.61$\times 10^{-3}$ & 60.2 & 4.25$\times 10^{-3}$ \\
2.1 & 19.0 & 2.53$\times 10^{-3}$ & 63.2 & 3.02$\times 10^{-3}$ \\
2.2 & 19.8 & 1.76$\times 10^{-3}$ & 66.2 & 2.10$\times 10^{-3}$ \\
2.3 & 20.8 & 1.24$\times 10^{-3}$ & 69.2 & 1.46$\times 10^{-3}$ \\
2.4 & 21.6 & 8.67$\times 10^{-4}$ & 72.2 & 1.01$\times 10^{-3}$ \\
2.5 & 22.6 & 6.07$\times 10^{-4}$ & 75.3 & 7.31$\times 10^{-4}$ \\
\hline
\end{tabular}
\end{center}
\end{table*}

\begin{figure}[htbp]
\centering
\includegraphics[width=0.49\textwidth]{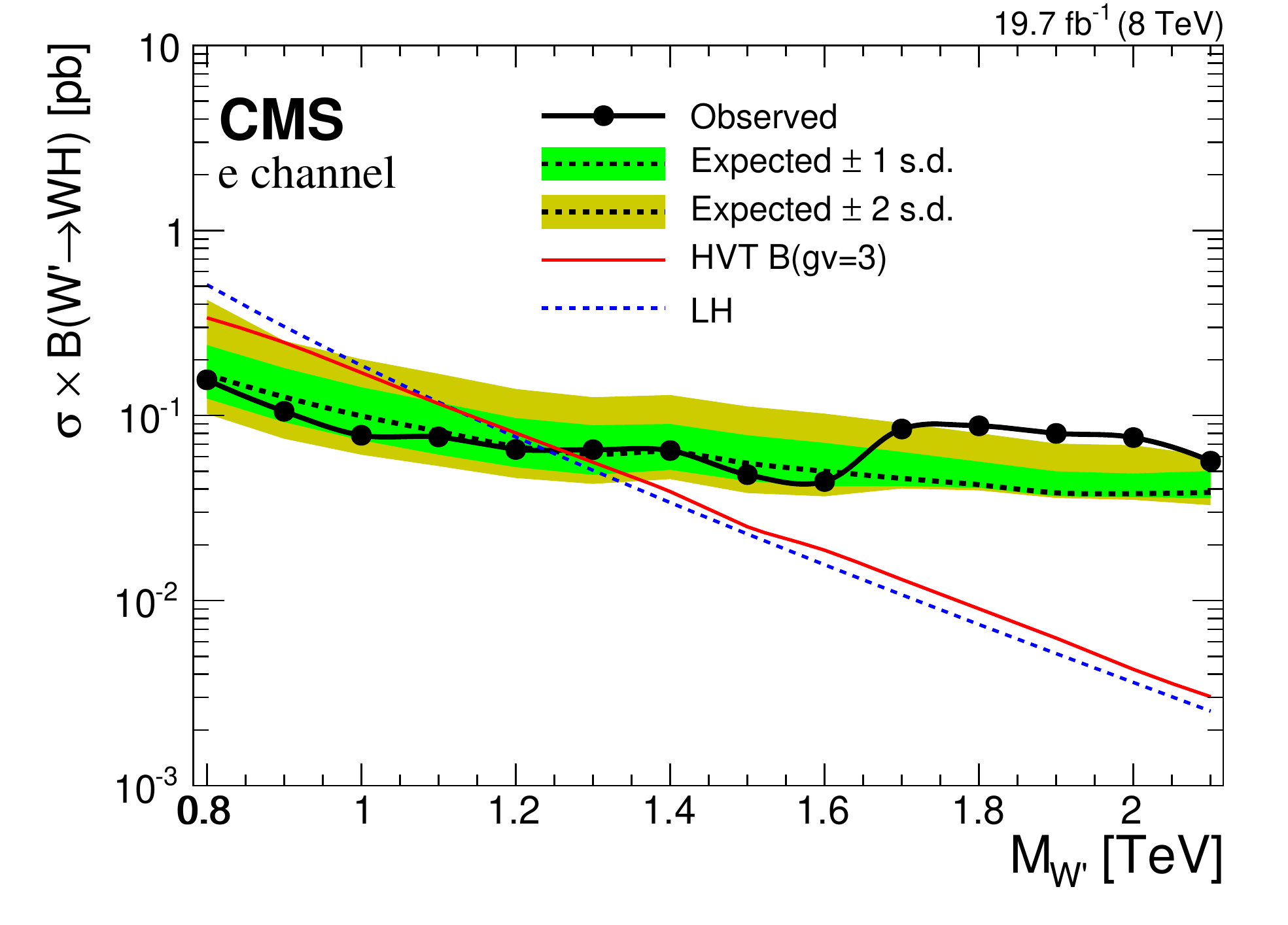}
\includegraphics[width=0.49\textwidth]{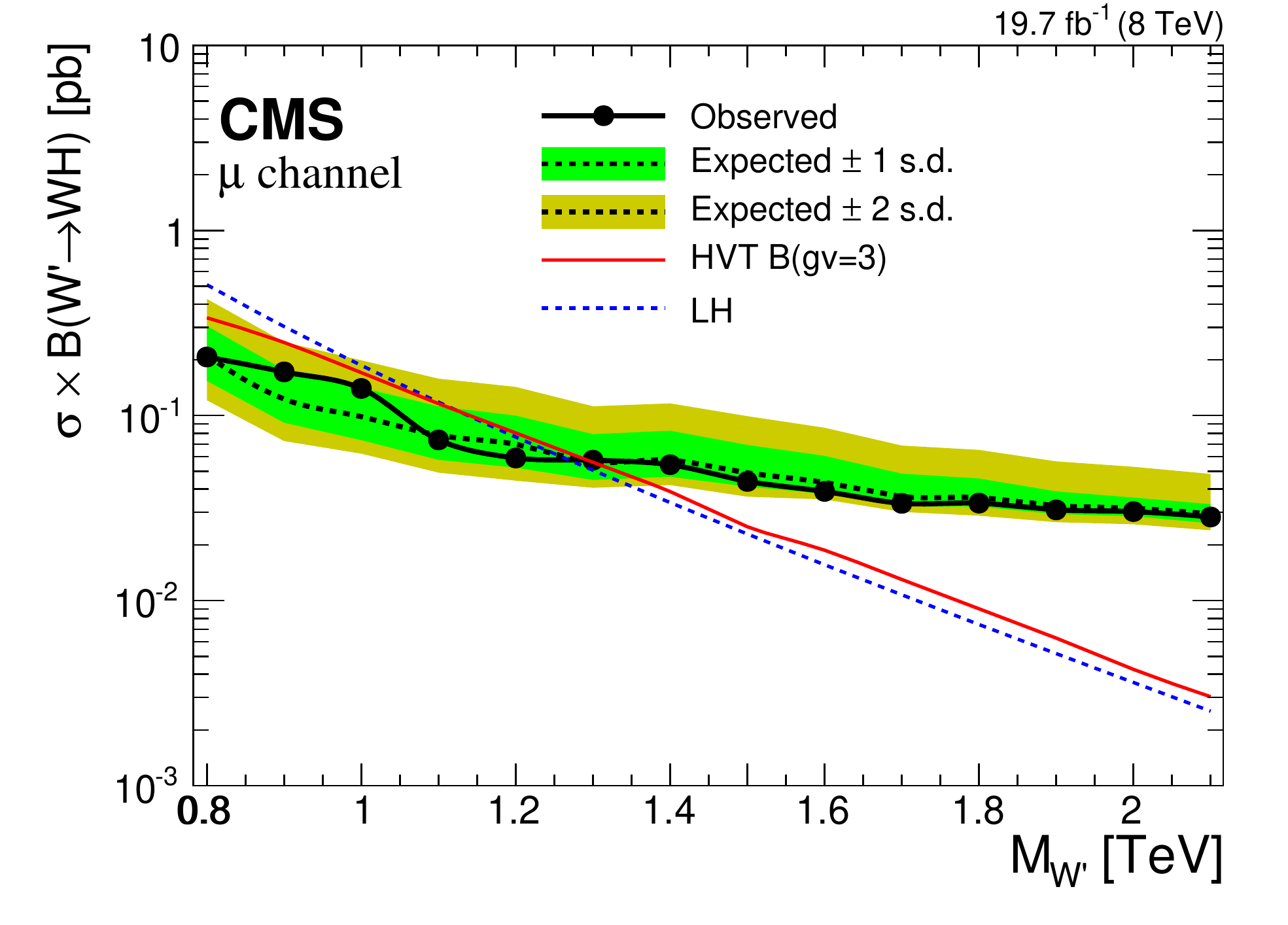}
\includegraphics[width=0.49\textwidth]{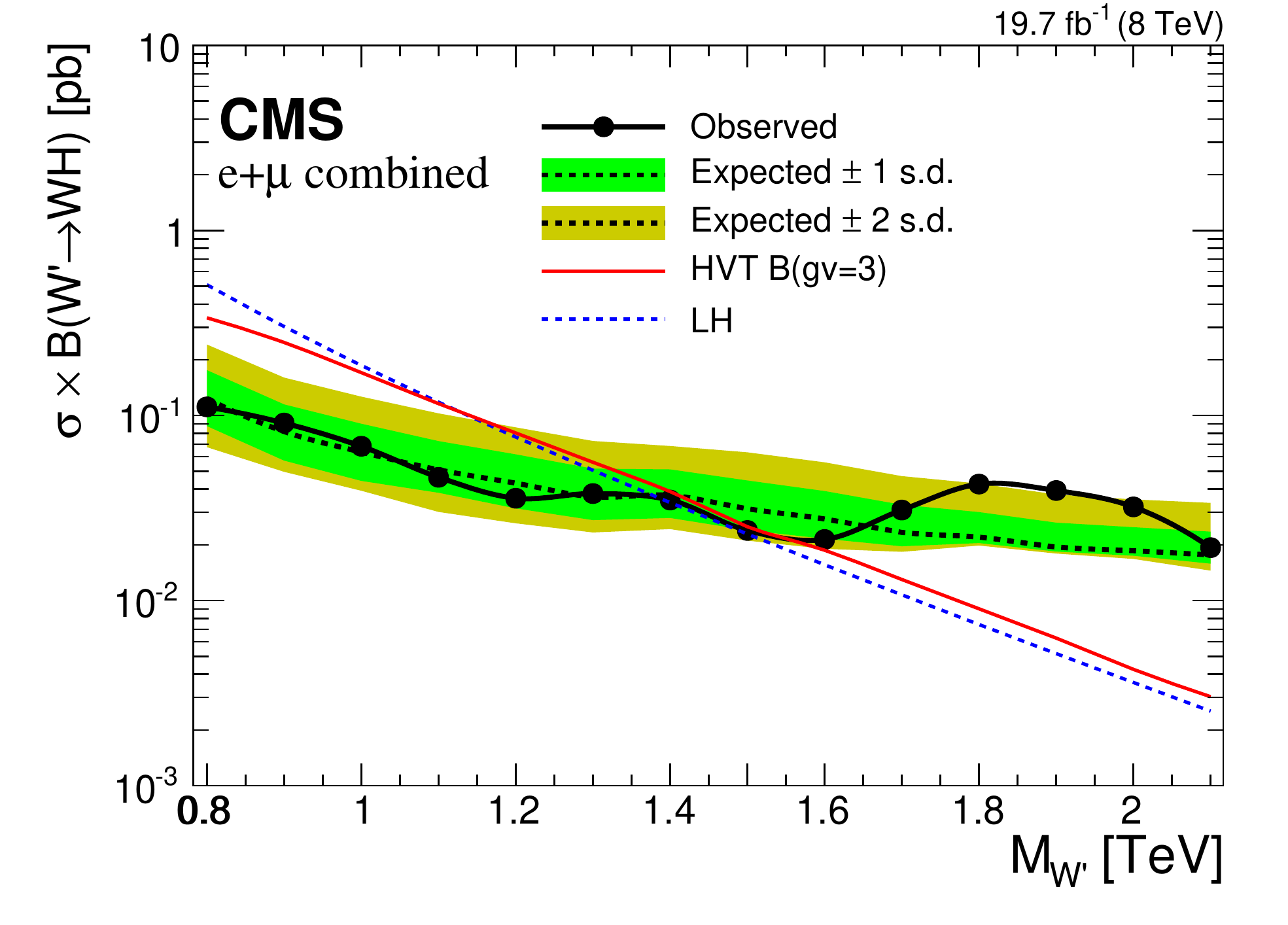}
\caption{
  Observed (solid) and expected (dashed) upper limits at 95\% CL on the
  product of the $\PWpr$ production cross section and the branching
  fraction of $\PWpr\to \PW\PH$ for electron (\cmsTopLeft) and muon (\cmsTopRight) channels,
  and the combination of the two channels (lower plot). The products of cross sections and branching fractions for $\PWpr$ production in the LH and HVT models are overlaid.
}
\label{fig:seperateLimits_FullCLs}
\end{figure}

Figure~\ref{fig:seperateLimits_FullCLs} shows the expected and observed exclusion limits at 95\% confidence level (CL)
on the product of the $\PWpr$ production cross section and the branching fraction of $\PWpr\to \PW\PH$
for the electron and muon channels separately, and for the combination of the two. For the combined channels, the observed and expected lower limits on the $\PWpr$ mass are 1.4\TeV in the
LH model and 1.5\TeV in the HVT model B.
For the electron (muon) channel, the observed and expected lower limits on the $\PWpr$ mass are 1.2\,(1.3)\TeV in the LH model and 1.3\,(1.3)\TeV in the HVT model B.

\subsection{Analysis combination}

\begin{figure}[htb]
\centering
     \includegraphics[width=0.49\textwidth]{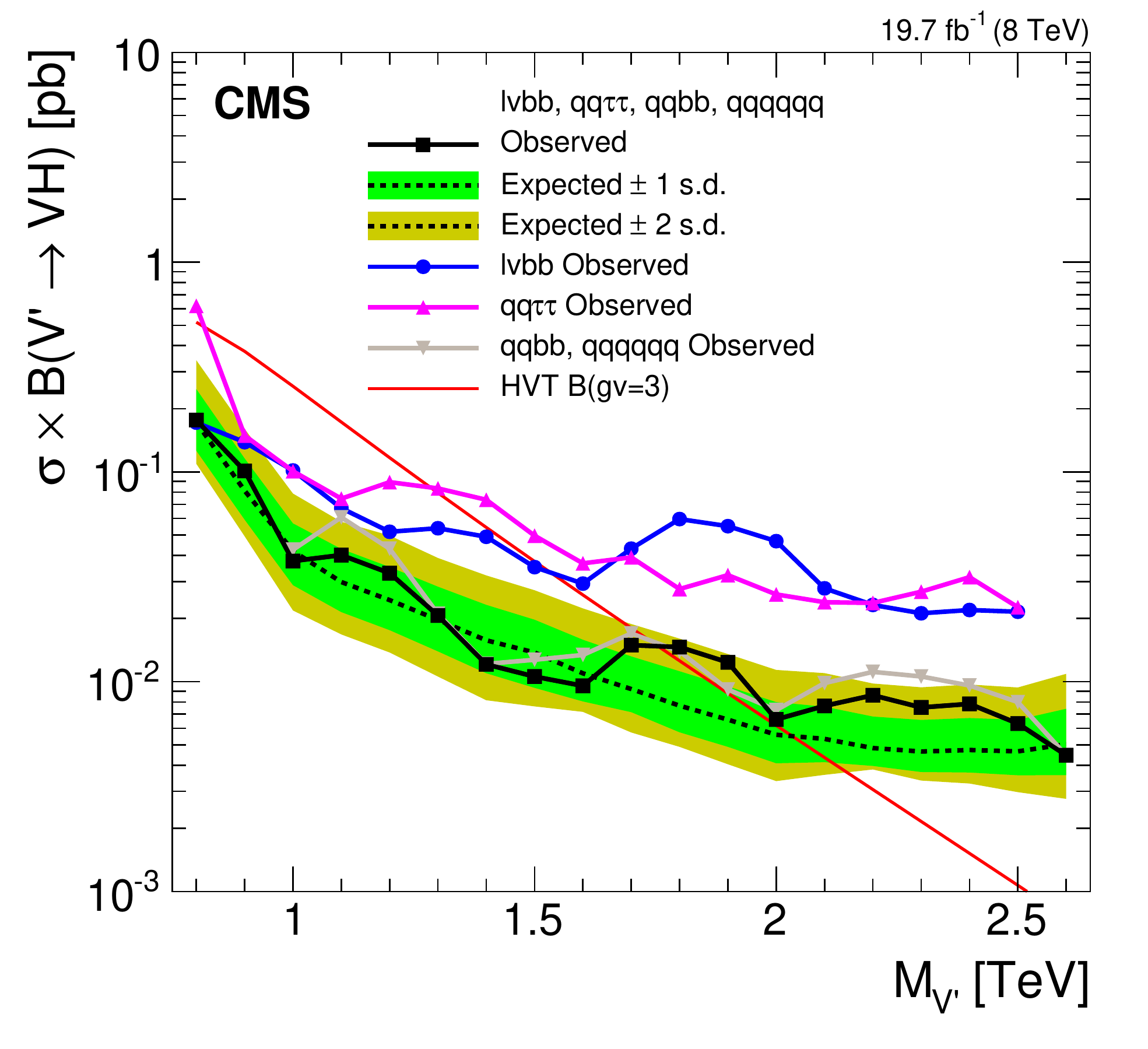}
\caption{
  Observed (full rectangles) and expected (dashed line) combined upper limits at 95\% CL on the sum of the $\PWpr$ and $\cPZpr$ production cross sections, weighted by their respective branching
  fraction of $\PWpr\to \PW\PH$ and $\cPZpr\to \cPZ\PH$.
  The cross section for the production of a $\PWpr$ and  $\cPZpr$ in the HVT model B, multiplied by its branching fraction for the relevant process, is overlaid.
  The observed limits of the three analyses entering the combination in the final states, $\ell\nu\bbbar$ (full circle),
  $\qqbar\tau^{+}\tau^{-}$~\cite{cms-HZ-tautaujet} (full triangle pointing up), and $\qqbar\bbbar/\qqbar\qqbar\qqbar$~\cite{Khachatryan:2015bma} (full triangle pointing down), are overlaid.
}
\label{fig:limitCombination}
\end{figure}

The limits obtained in this analysis can be combined with two previous results~\cite{cms-HZ-tautaujet,Khachatryan:2015bma},
setting limits on the sum of $\PWpr\to \PW\PH$ and $\cPZpr\to \cPZ\PH$ production in the context of the HVT model.
The search for $\PWpr/\PZpr \to \PW\PH/\Z\PH \to \PQq'\PAQq\bbbar/\qqbar\qqbar\qqbar$~\cite{Khachatryan:2015bma} reports
limits in the context of the HVT model that can be directly used in the combination.
However, while an asymptotic approximation of the \CLs procedure was used in the original paper, for the combination the limit is re-evaluated with the \CLs procedure reported above.
The search for $\PZpr \to \Z\PH \to \qqbar \tau^{+}\tau^{-}$~\cite{cms-HZ-tautaujet},
does not report limits in the context of a $\PWpr$ resonance.
However, since it is also sensitive to a signal from $\PWpr \to \PW\PH \to \PQq'\PAQq\tau^{+}\tau^{-}$
with an efficiency of about 5\% less than for the $\PZpr$ signal, it was reinterpreted for the purpose of the combination.
The results of the combination are shown in Fig.~\ref{fig:limitCombination}.
The limit on the mass of the $\PWpr/\PZpr$ is slightly improved to 1.8\TeV compared to the most stringent result
reported by the $\PWpr/\PZpr \to \PW\PH/\Z\PH \to \PQq^\prime\PAQq\bbbar/\qqbar\qqbar\qqbar$ search.

\begin{figure}[htb]
\centering
     \includegraphics[width=\cmsFigWidth]{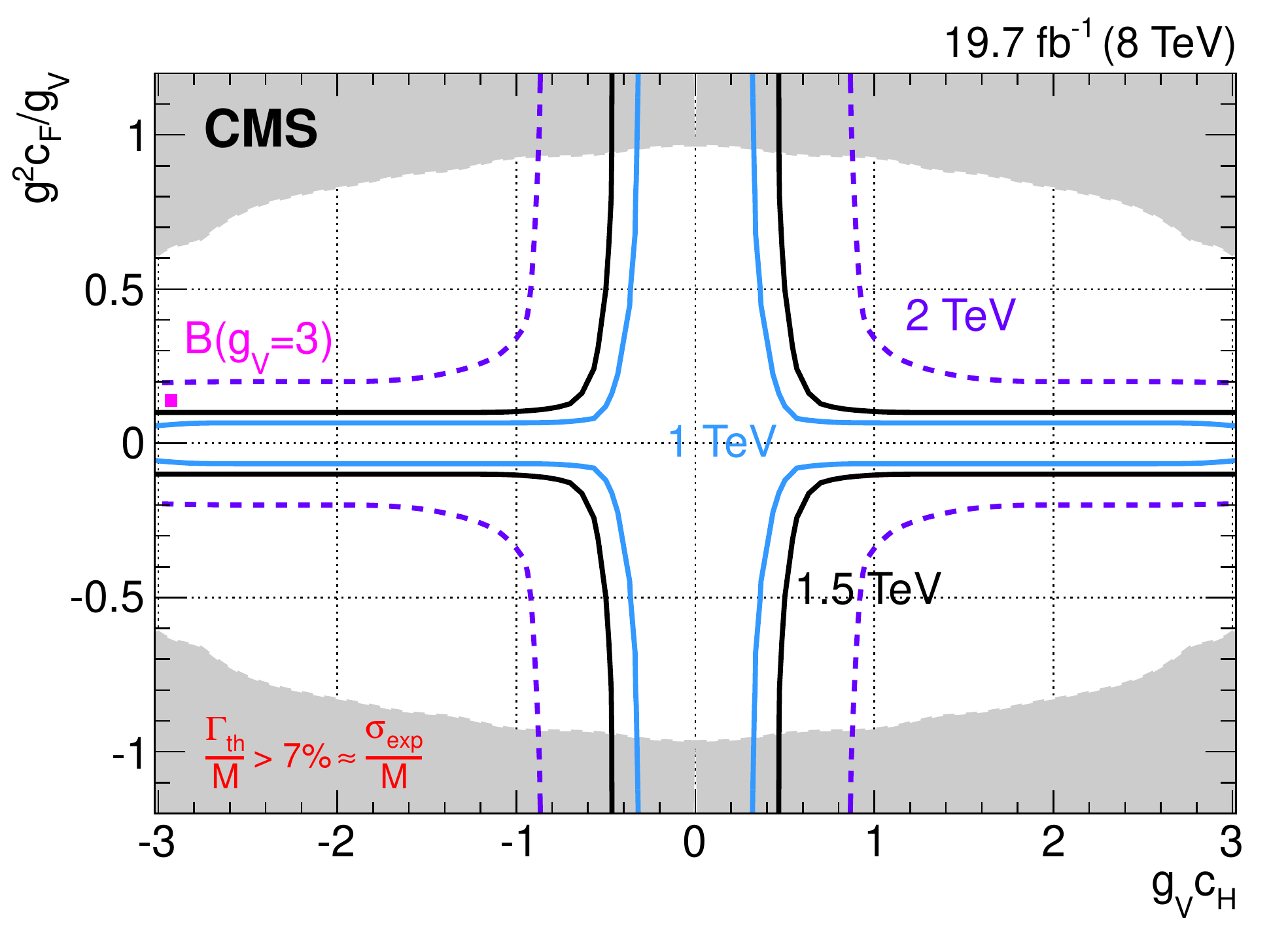}
\caption{
Exclusion regions in the plane of the HVT-model couplings ($g_\Vo c_\PH$, $g^{2}c_\mathrm{F}/g_\Vo$) for three
resonance masses, 1, 1.5, and 2\TeV, where $g$ denotes the weak gauge coupling. The point B of the benchmark model used in the analysis is also shown.
The boundaries of the regions outside these lines are excluded by this search are indicated by the solid and dashed lines (region outside these lines is excluded).
The areas indicated by the solid shading correspond to regions
where the resonance width is predicted to be more than 7\% of the resonance mass and the narrow-resonance assumption is not satisfied.
}
\label{fig:HVTCouplings}
\end{figure}

In Fig.~\ref{fig:HVTCouplings}, a scan of the coupling parameters and the corresponding observed 95\% CL exclusion
contours in the HVT model from the combination of the analyses are shown.
The parameters are defined as $g_\Vo c_\PH$ and $g^{2}c_{F}/g_\Vo$,
related to the coupling strengths of the new resonance to the Higgs boson and to fermions.
The range of the scan is limited by the assumption that the new resonance is narrow.
A contour is overlaid, representing the region where the theoretical width is larger than the experimental resolution of
the searches, and hence where the narrow-resonance assumption is not satisfied. This contour is defined by
a predicted resonance width of 7\%, corresponding to the largest resonance mass resolution of the considered searches.

\section{Summary}

A search has been presented for new resonances decaying into WH, in which the W boson decays into $\ell\nu$
with $\ell= \Pe$, $\mu$ and the Higgs boson decays to a pair of bottom quarks.
Each event is reconstructed as a leptonic W boson candidate
recoiling against a jet with mass compatible with the Higgs boson mass.
A specialized b tagging method for Lorentz-boosted Higgs bosons is used to further reduce the
background from multijet processes.
No excess of events above the standard model prediction is observed in the muon channel, while an excess with a local significance of 2.9 standard deviations is observed in the
electron channel near $\mWH\approx 1.8\TeV$. The results are statistically compatible with the standard model within 2 standard
deviations.
In the context of the little Higgs and the heavy vector triplet models,
upper limits at 95\% confidence level are set on the $\PWpr$ production cross section
in a range from 100 to 10\unit{fb} for masses between 0.8 and 2.5\TeV, respectively.
Within the little Higgs model, a lower limit on the $\PWpr$ mass of 1.4\TeV has been set. A heavy vector triplet model that mimics the properties of composite Higgs models has been excluded up to a $\PWpr$ mass of 1.5\TeV. In this latter context,
the results have been combined with related searches, improving the lower limit up to $\approx$1.8\TeV. This combined limit is the most restrictive to date for $\PWpr$ decays to a pair of standard model bosons.

\begin{acknowledgments}
We congratulate our colleagues in the CERN accelerator departments for the excellent performance of the LHC and thank the technical and administrative staffs at CERN and at other CMS institutes for their contributions to the success of the CMS effort. In addition, we gratefully acknowledge the computing centres and personnel of the Worldwide LHC Computing Grid for delivering so effectively the computing infrastructure essential to our analyses. Finally, we acknowledge the enduring support for the construction and operation of the LHC and the CMS detector provided by the following funding agencies: BMWFW and FWF (Austria); FNRS and FWO (Belgium); CNPq, CAPES, FAPERJ, and FAPESP (Brazil); MES (Bulgaria); CERN; CAS, MoST, and NSFC (China); COLCIENCIAS (Colombia); MSES and CSF (Croatia); RPF (Cyprus); MoER, ERC IUT and ERDF (Estonia); Academy of Finland, MEC, and HIP (Finland); CEA and CNRS/IN2P3 (France); BMBF, DFG, and HGF (Germany); GSRT (Greece); OTKA and NIH (Hungary); DAE and DST (India); IPM (Iran); SFI (Ireland); INFN (Italy); MSIP and NRF (Republic of Korea); LAS (Lithuania); MOE and UM (Malaysia); CINVESTAV, CONACYT, SEP, and UASLP-FAI (Mexico); MBIE (New Zealand); PAEC (Pakistan); MSHE and NSC (Poland); FCT (Portugal); JINR (Dubna); MON, RosAtom, RAS and RFBR (Russia); MESTD (Serbia); SEIDI and CPAN (Spain); Swiss Funding Agencies (Switzerland); MST (Taipei); ThEPCenter, IPST, STAR and NSTDA (Thailand); TUBITAK and TAEK (Turkey); NASU and SFFR (Ukraine); STFC (United Kingdom); DOE and NSF (USA).

Individuals have received support from the Marie-Curie programme and the European Research Council and EPLANET (European Union); the Leventis Foundation; the A. P. Sloan Foundation; the Alexander von Humboldt Foundation; the Belgian Federal Science Policy Office; the Fonds pour la Formation \`a la Recherche dans l'Industrie et dans l'Agriculture (FRIA-Belgium); the Agentschap voor Innovatie door Wetenschap en Technologie (IWT-Belgium); the Ministry of Education, Youth and Sports (MEYS) of the Czech Republic; the Council of Science and Industrial Research, India; the HOMING PLUS programme of the Foundation for Polish Science, cofinanced from European Union, Regional Development Fund; the OPUS programme of the National Science Center (Poland); the Compagnia di San Paolo (Torino); MIUR project 20108T4XTM (Italy); the Thalis and Aristeia programmes cofinanced by EU-ESF and the Greek NSRF; the National Priorities Research Program by Qatar National Research Fund; the Rachadapisek Sompot Fund for Postdoctoral Fellowship, Chulalongkorn University (Thailand); the Chulalongkorn Academic into Its 2nd Century Project Advancement Project (Thailand); and the Welch Foundation, contract C-1845.\end{acknowledgments}
\bibliography{auto_generated}

\cleardoublepage \appendix\section{The CMS Collaboration \label{app:collab}}\begin{sloppypar}\hyphenpenalty=5000\widowpenalty=500\clubpenalty=5000\textbf{Yerevan Physics Institute,  Yerevan,  Armenia}\\*[0pt]
V.~Khachatryan, A.M.~Sirunyan, A.~Tumasyan
\vskip\cmsinstskip
\textbf{Institut f\"{u}r Hochenergiephysik der OeAW,  Wien,  Austria}\\*[0pt]
W.~Adam, E.~Asilar, T.~Bergauer, J.~Brandstetter, E.~Brondolin, M.~Dragicevic, J.~Er\"{o}, M.~Flechl, M.~Friedl, R.~Fr\"{u}hwirth\cmsAuthorMark{1}, V.M.~Ghete, C.~Hartl, N.~H\"{o}rmann, J.~Hrubec, M.~Jeitler\cmsAuthorMark{1}, V.~Kn\"{u}nz, A.~K\"{o}nig, M.~Krammer\cmsAuthorMark{1}, I.~Kr\"{a}tschmer, D.~Liko, T.~Matsushita, I.~Mikulec, D.~Rabady\cmsAuthorMark{2}, B.~Rahbaran, H.~Rohringer, J.~Schieck\cmsAuthorMark{1}, R.~Sch\"{o}fbeck, J.~Strauss, W.~Treberer-Treberspurg, W.~Waltenberger, C.-E.~Wulz\cmsAuthorMark{1}
% ...
\vskip\cmsinstskip
% ...
1:~~Also at Vienna University of Technology, Vienna, Austria\\
2:~~Also at CERN, European Organization for Nuclear Research, Geneva, Switzerland\\
% ...

\end{sloppypar}
\end{document}